\documentclass[fleqn,usenatbib]{mnras}

\usepackage{newtxtext,newtxmath}

\usepackage[T1]{fontenc}

\DeclareRobustCommand{\VAN}[3]{#2}
\let\VANthebibliography\thebibliography
\def\thebibliography{\DeclareRobustCommand{\VAN}[3]{##3}\VANthebibliography}

\usepackage{graphicx} 
\usepackage[dvipsnames]{xcolor}  
\usepackage{hyperref}

\usepackage{graphicx}	
\usepackage{amsmath}	

\title[Needlet Karhunen-Loève (NKL)]{Needlet Karhunen-Loève (NKL): A Method For Cleaning Foregrounds From 21\,cm Intensity Maps}

\author[Podczerwinski and Timbie]{
John Podczerwinski$^{1}$\thanks{E-mail: jpodczerwins@wisc.edu} and 
Peter T. Timbie$^{1} \thanks{E-mail: pttimbie@wisc.edu}$
\\
$^{1}$Department of Physics, University of Wisconsin - Madison\\
1150 University Avenue, Madison, WI 53706\\
}

\date{Accepted XXX. Received YYY; in original form ZZZ}

\pubyear{2023}

\begin{document}
\label{firstpage}
\pagerange{\pageref{firstpage}--\pageref{lastpage}}
\maketitle

\begin{abstract}
This paper introduces a technique called NKL, which cleans both polarized and unpolarized foregrounds from HI intensity maps by applying a Karhunen-Loève transform on the needlet coefficients. In NKL, one takes advantage of correlations not only along the line of sight, but also between different angular regions, referred to as ``chunks". This provides a distinct advantage over many of the standard techniques applied to map-space that one finds in the literature, which do not consider such spatial correlations. Moreover, the NKL technique does not require any priors on the nature of the foregrounds, which is important when considering polarized foregrounds. We also introduce a modified version of GNILC, referred to as MGNILC, which incorporates an approximation of the foregrounds to improve performance. The NKL and MGNILC techniques are tested on simulated maps which include polarized foregrounds. Their performance is compared to the GNILC, GMCA, ICA and PCA techniques. Two separate tests were performed. One at $1.84 < z < 2.55$ and the other at $0.31 < z < 0.45$. NKL was found to provide the best performance in both tests, providing a factor of 10 to 50 improvement over GNILC at $k < 0.1\,{\rm hMpc^{-1}}$ in the higher redshift case and $k < 0.03 \,{\rm hMpc^{-1}}$ in the lower redshift case. However, none of the methods were found to recover the power spectrum satisfactorily at all BAO scales. 
\end{abstract}







\section{Introduction}\label{sec:intro}

One of the main endeavors of observational cosmology is to measure statistical properties of the spatial distribution of matter in the Universe. Such measurements are of interest as they could provide us with information about dark energy, inflationary physics, the growth of structure, early star formation and more. The distribution of matter in the Universe is traced by neutral hydrogen (HI). Using 21\,cm emission or absorption to measure these density fluctuations as a function of redshift is a technique referred to as 21\,cm (or HI) intensity mapping (IM) \cite{Bharadwaj_2001} \cite{battye} \cite{Madau_1997}. Most IM instruments are radio interferometers; some notable examples are the Canadian Hydrogen Intensity Mapping Experiment (CHIME) \cite{CHIME}, Tianlai \cite{Tianlai}, the Hydrogen Epoch of Reionization Array (HERA) \cite{HERA} and the Murchison Widefield Array (MWA) \cite{MWA}. The antenna elements that make up these interferometers can take several forms, including parabolic-dishes (Tianlai, CHIME, HERA), cylindrical reflectors (CHIME, Tianlai) or phased arrays (MWA).  Single-dish instruments are also used, with some examples being the Five hundred meter Aperture Spherical Telescope (FAST) \cite{FAST}, the Greenbank Telescope (GBT) \cite{greenbank} and the More Karoo Array Telescope (MeerKAT) \cite{meerkat}. FAST and GBT both consist of a single, very large dish. On the other hand, MeerKAT would average signals from 64 smaller ($13.5$\,m diameter) dishes. MeerKAT can also operate as an interferometer.

Measurement of the 21\,cm line  would provide valuable  cosmological information. At low redshifts ($z \leq 6$), IM would serve as a complement to galaxy redshift surveys. At such redshifts, cosmologists would be particularly interested in measuring Baryon Acoustic Oscillations (BAO). The BAO would serve as a standard ruler, providing information about the expansion of the Universe and dark energy.  Measurements during the epoch of reionization $6 < z < 20$ would provide information about the formation of the earliest stars. Lastly, measurements during the cosmic dark ages ($20 < z < 1100$), before the formation of the first luminous objects, would provide insights into the physics of inflation. It should be noted that no other probe aside from IM is capable of mapping the cosmic dark ages. 

Although promising, IM is still a young technique, with its systematic effects and calibration requirements still being understood. In particular, foregrounds from Galactic synchrotron emission are an unsolved challenge for IM experiments. In the case of Galactic synchrotron emission, these foregrounds are up to a factor of $10^{5}$ brighter than the HI signal.

Luckily, the unpolarized component of these foregrounds is expected to be spectrally smooth, while the HI signal is expected to have a high level of chromaticity. In other words, the foregrounds are confined to a relatively small subsection of delay-space \cite{Datta_2010}. In this context, delay refers to the Fourier dual of spectral frequency. This delay-space quarantining aids in the avoidance and removal of the foregrounds. Unluckily, this region of delay space will correspond to large spatial scales, which are important for measuring the BAO. 

One must also consider the polarized component of Galactic synchrotron radiation. This component is affected by Faraday rotation in the Galaxy, introducing higher chromaticity than the unpolarized component. Moreover, the chromaticity of this component is expected to vary with line of sight direction. In particular, lines of sight closer to the Galactic plane will suffer from more severe chromaticity due to stronger Faraday rotation. Moreover, the power of the Galactic synchrotron radiation (both polarized and unpolarized components) relative to the signal is expected to vary with angular scale \cite{Alonso_2014}, with the largest scales having the worst contamination. In principle, these polarized foregrounds could be avoided altogether as long as the  beams of the telescope have rotational symmetry and low cross-polar levels. However, such refined beams are not achievable in practice, and one must contend with some amount of polarization leakage. This leakage is expected to be on the percent level for IM experiments \cite{Moore_2013} \cite{Moore_2017}. 

Over the years, many foreground removal methods have been proposed. A helpful review of many of these may be found in \citet{Liu_Shaw}. Methods that require no priors on the HI, noise or foregrounds are described as ``blind", and those requiring priors are described as ``non-blind". One will also find methods being tested both on raw visibilities or on maps sythesized from data. However, unlike blind and non-blind, there is not a clear distinction here. Some methods may be applicable both in map-space and in visibility-space.

Principal Component Analysis (PCA), Independent Component Analysis (ICA) \cite{Chapman_ica} and Generalized Morphological Component Analysis (GMCA) \cite{Chapman_gmca} are the prominent blind methods that one finds being applied to maps. In the literature, one can also find non-blind methods being applied in map-space. Examples include Generalized Needlet Internal Linear Combination (GNILC) \cite{Olivari_2015}, Gaussian Process Regression (GPR) \cite{mertens_2018} and the ``semi-blind" Singular Value Decomposition (SVD) method proposed in \citet{Zuo_2023}. On the other hand, tests of cleaning methods on visibilities are less commonly found in the literature. One blind method that can be used in visibility space is  ``foreground avoidance". In this method, one exploits the fact that the foregrounds tend to be confined to a region of certain region $k$-space, referred to as ``the wedge" \cite{delay_spec}. Power spectrum modes within this region of k-space are then excluded from the analysis. Non-blind methods have also been tested in visibility space. The beam projection plus Karhunen-Loève transform proposed in \citet{Shaw_2015} and the GPR method used in \citet{Soares_2021} and \citet{mertens_2018} are examples we have found in the literature.

However, for the case of post-EOR surveys, such avoidance methods come at the cost of losing valuable information about Baryon Acoustic Oscillations (BAO). The BAO are located roughly in the wavenumber range $0.03 \leq k \leq 0.4 \,{\rm hMpc^{-1}}$ \cite{Bull_2015}. Much information about large length scales such as these would be lost when taking a foreground avoidance approach.

So far, when real data is considered, the foregrounds have usually been handled in a conservative way. For instance, in \citet{paul2023detection}, the first detection of HI using IM without cross-correlating with galaxies, the analysts chose to use foreground avoidance rather than removal. Foreground avoidance was also used in \citet{CHIME_2023}. In this work, the authors cross-correlated data from the CHIME telescope with galaxies and quasars from eBOSS. In this work, it was found that the foreground avoidance method excluded length scales important for characterizing the BAO. The authors \citet{Wolz_2021} present results from cross-correlating GBT data with the eBOSS survey. In this case, the authors opted to use FastICA paired with a transfer function to compensate for signal loss. In \citet{Cunnington_2023}, the authors cross-correlated maps from MeerKAT with WiggleZ galaxies. The authors of this paper cleaned their data with PCA and used a transfer function to compensate for signal loss. The authors of \citet{anderson_2018} cross-correlated maps from the Parkes telescope with galaxies from the 2dF galaxy survey. These authors correlated maps from different seasons to reduce noise bias, and applied a transfer function to compensate for signal loss.

One can also find numerous papers in the literature testing these various methods on simulated data. Such tests are usually conducted at lower redshifts ($z \lesssim 0.6$) and take into account beam effects by convolving with a Gaussian profile of appropriate width. In \citet{Carucci_2020}, tests were conducted using simulations that assumed a telescope similar to MeerKAT operating in single dish mode at low redshifts ($0.09 \leq z \leq 0.58$) and surveying the full sky. This test included contributions from polarized foregrounds simulated using the Cosmological Realizations for Intensity Mapping Experiments (CRIME) software package \cite{Alonso_2014}. In this instance, GMCA recovered the angular power spectrum of the HI plus noise with errors of $5\%$ down to $\ell \approx 25$. Moreover, they found that GMCA provided lower errors than ICA \cite{Carucci_2020}. In \citet{Olivari_2015}, GNILC was tested at low redshifts ($0.13 \leq z \leq 0.48$) in a survey covering half of the sky. These simulated maps included no polarized foregrounds. In this case, GNILC was found to recover the angular power spectrum of the HI to within $6 \%$ error down to $\ell = 30$. So, like GMCA, GNILC also seemed to struggle at large angular scales. Recently, GNILC, GMCA and ICA were all tested on simulations of the BINGO experiment \cite{marins2022foreground}. In this test, all of the techniques were found to provide statistically equivalent results. In \citet{Soares_2021}, GPR was tested on simulated maps at low redshifts ($0.18 < z < 0.58$). The simulated maps included polarized foregrounds generated using CRIME. These tests were conducted on Stripe82 and a $3000 \,{\rm deg^{2}}$ region at the South Celestial Pole (SCP). In these tests, GPR and PCA provided similar performance, recovering the power spectrum with errors around $10 \%$ for all spatial scales considered.

A notable cleaning method applied in visibility-space is the combined beam projection and Karhunen-Loève (KL) transform proposed and tested in \citet{Shaw_2015}. This test assumed a simplified version of the CHIME instrument and was conducted at higher redshifts ($1.84 < z < 2.55$) where polarized foregrounds are more severe. The test used simulated maps produced by the Cosmology in the Radio Band (CORA) software package. This package makes different assumptions than CRIME, resulting in less severe chromaticity in the polarized foregrounds. In this paper, instrumental effects were accounted for in a more realistic  way, by generating visibilities from maps using simulated beams from cylinder telescopes. This method was found to effectively recover the HI power spectrum down to $k \approx 0.02\,{\rm hMpc^{-1}}$. This result appears quite promising, since the BAO would be recovered at all length scales. However, the weakness of this approach is that it requires a highly accurate beam model along with priors on the unpolarized foregrounds, HI and noise. It was found that this approach breaks down when main beam uncertainties exceed $0.1\%$. In addition, amplifier gains must be known to better than 1\% within each minute. Such accurate beam and gain calibration pose a significant challenge and may not be possible in practice.

In summary, published map-space tests of cleaning methods on simulated polarized foregrounds have only been conducted at low redshifts. Moreover, these papers usually take beam effects into account in a simple way. The only exception we found is in the work of \citet{Hothi_2020}, who performed tests on maps generated from visibilities. Such tests are informative, but not the whole story since these polarized foregrounds will become more severe as redshift increases. On the other hand, Shaw's method was found to work extremely well when faced with polarized foregrounds at higher redshifts, but required precise knowledge of the beam. It is reasonable to imagine that the previously described methods used in map-space might be more robust to beam mis-calibration than Shaw's method. There are two reasons for this. One is that the beam projection part of beam projection/KL approach removes the polarized foregrounds by projecting onto the null-space of the polarized beam matrix. This step may not work when the beams are not well understood. Another reason is that the KL part of the SVD/KL cleaning requires one to have priors on how the unpolarized foregrounds will contribute to the total visibility. On the other hand, the map-space applicable methods described earlier in this paper (aside from the ``semi-blind" SVD method from \citet{Zuo_2023})  either do not require a foreground model, or, in the case of GPR, estimate one from the data. 

Although they may end up being more robust to calibration issues than the visibility-space beam projection/KL method, the available methods usable in map-space are missing certain strengths of the beam projection/KL approach. One particular strength of beam projection/KL is that it takes full advantage of the available priors. In particular, during the KL step, one considers correlations between all baselines in the telescope. In this way, one is exploiting correlations in both frequency and spherical harmonics (different baselines are sensitive to different l-modes). This is much different from commonly used map-space methods, such as PCA, GNILC, ICA, and GPR, which, as employed so far in the literature, only consider correlations along the line of sight. Although there are some exceptions, such as the commonly used GMCA method does incorporate angular correlations by enforcing sparsity of components in the wavelet domain. Another exception to this is the ``semi-blind" SVD method. This method provides excellent recovery of the HI signal, but, like Shaw's method, suffers from requiring priors on the foregrounds present in the maps.  

However, one must also keep in mind that working in map-space presents certain advantages not available in visibility space. For one, map-space allows for specific pixels to be selected. As such, one can mask out pixels with particularly strong foregrounds or artifacts. Moreover, one can perform a spherical-harmonic or spherical wavelet transform of the data, allowing for precise separation by angular scale. Such operations are not possible in visibility space. The closest thing one could do is separate the visibilities into m-modes, where the m refers to the azimuthal m found in spherical harmonics \cite{Shaw_2015}. Such freedom could be useful when cleaning foregrounds from maps, due to their dependence on line of sight direction and angular scale.  

Based on this review of the literature, there several tasks that ought to be performed. 
\begin{enumerate} \label{todolist}
  \item Create a foreground cleaning method that is usable in map space and uses both frequency and angular correlations. This method should not require a prior on the foregrounds.
\item Test the available map-based methods on polarized foregrounds at redshifts $z \gtrsim 0.6$.
  
    \item Test the robustness of various map-based methods against beam mis-calibration. 
\end{enumerate}

We make progress on task (i) by introducing the Needlet Karhunen-Loève (NKL) method of foreground removal. In NKL, different sections of pixel/spherical harmonic space are separated via a needlet transform. The needlet coefficients are then cleaned using a Karhunen-Loève transform that exploits both angular and frequency correlations. This is different from most other map-based methods, which only consider frequency correlations. Moreover, this is all done without needing any priors on the foregrounds present in the map. Not requiring a foreground prior is quite useful, as polarized foregrounds are not well understood. Moreover, NKL cleans the foregrounds differently depending on location and angular scale. This is also a desirable feature, as the foreground brightness and chromaticity are expected to vary with line of sight and angular scale.

In this paper, we also make some progress on task (ii) of the list by testing GNILC, GMCA, ICA and PCA at redshifts ($1.84 < z < 2.55)$. PCA, ICA and GMCA were chosen as these are all prominent in the literature. GNILC was chosen since it has similarities to the NKL method proposed in this paper.

We leave task (iii) for future work.

In Section~\ref{section:bigpicture}, we describe the process of foreground cleaning in an abstract way and provide additional motivation for the NKL technique. In Section~\ref{section:nkl}, we provide a detailed description of the NKL technique. In this Section, we also introduce Modified GNILC (MGNILC), a slightly modified version of GNILC which uses a foreground approximation acquired from the data. In Section~\ref{sec:sim_maps}, we describe the maps we used to perform our tests. In Section~\ref{section:tests}, we present results acquired by testing GNILC, ICA, PCA, GMCA and NKL on simulated maps. We then summarize our results and conclude in Section~\ref{section:conclusion}.

\section{Foreground Removal Techniques} \label{section:bigpicture}
In this paper, we will compare the performance of the proposed NKL technique with that of other techniques commonly found in the literature. In particular, we will be considering Principal Component Analysis (PCA) \cite{Cunnington_PCA}, Independent Component Analysis (ICA)  \cite{Wolz_2014} , Generalized Morphological Component Analysis (GMCA) \cite{Carucci_2020}, and Generalized Needlet Internal Linear Combination (GNILC) \cite{Olivari_2015}. Although not quite obvious at first, these techniques are in fact very similar. In particular, all techniques tested in this paper will model the foregrounds (or signal plus noise in the case of GNILC) as a mixture of template maps. 
\subsection{Foreground Removal Using Templates} \label{subsection:fgremoval}
All techniques considered in this paper begin by assuming that the maps produced by an IM experiment are given by 
\begin{equation}
    \mathbf{X} = \mathbf{f} + \mathbf{h} + \mathbf{n}.
\end{equation}
In this formula, $\mathbf{X}$ is a $n_{ch} \times n_{p}$ matrix, where $n_{ch}$ is the number of frequency channels and $n_{p}$ is the number of pixels in each map. Moreover, $\mathbf{f}$ represents the foregrounds, $\mathbf{h}$ represents the HI signal and $\mathbf{n}$ represents the noise. In this paper we will use bold font to denote matrices. We will also define 
\begin{equation} \label{eq:sdef}
    \mathbf{s} = \mathbf{h} + \mathbf{n}.
\end{equation}

It is then assumed that the foregrounds can be expressed as a mixture of templates:
\begin{equation}
    \mathbf{f} = \mathbf{A}\mathbf{S}.
\end{equation}
In this formula, $\mathbf{S}$ is a $n_{t} \times n_{p}$ matrix of templates, where $n_{t}$ is the number of templates and  $\mathbf{A}$ is a $n_{ch} \times n_{t}$ `mixing matrix' which encodes how the templates are combined at each frequency channel. The foreground removal process then becomes a matter of fitting $\mathbf{A}\mathbf{S}$ to $\mathbf{X}$, subject to some sort of regularization. 

In PCA, one seeks to find templates that capture as much variation in the data as possible. In particular, these templates are really just the dot product of the data with the eigenvectors corresponding to the largest $n_{t}$ eigenvalues of the covariance matrix. In this context, the covariance matrix is usually generated assuming that the brightness of the sky along each line of sight were independently drawn from some distribution. This assumption is not true, but is close enough to reality that PCA can still provide reasonable results. In the context of 21\,cm intensity mapping, one usually estimates the covariance from the data as 
\begin{equation}
   \hat{\mathbf{C}} =  \frac{1}{n_{p}} (\mathbf{X}-\overline{\mathbf{X}}) (\mathbf{X}-\overline{\mathbf{X}})^{T}, 
\label{eq:pca_cov}
\end{equation}
where row $i$ of  $\overline{\mathbf{X}}$ is the average of row $i$ of  $\mathbf{X}$. In this paper, we will use hats to denote covariance estimates. On the other hand, we will use matrices without hats to denote true covariances. In this case, $\hat{C}$ is a $n_{ch} \times n_{ch}$ matrix. 
In ICA, one seeks templates that are statistically independent. In GMCA, one assumes that the templates $\mathbf{S}$ ought to be sparse in some wavelet domain. ILC methods such as GNILC seek to find a filter that has a unit response to $\mathbf{h} + \mathbf{n}$ while minimizing the variance of the residuals \cite{Olivari_2015}. Readers can refer to \citet{marins2022foreground} for a more rigorous description of how these standard techniques are regularized.

Both Shaw's work and the Needlet Karhunen-Loève (NKL) technique described in this paper make use of the Karhunen-Loève (KL) transform \cite{Tegmark_1997} for foreground cleaning. The KL transform uses the covariance matrices of both the foregrounds and signal to clean the data. These covariances either come from priors or are estimated from the data itself. For NKL, we estimate the foreground covariance from the data and get the HI and noise covariances from priors. Let $\mathbf{C}_{FG}$ be the foreground covariance and $\mathbf{C}_{S}$ be the signal covariance. For now, let's consider only frequency correlations, resulting in matrices of size $n_{ch} \times n_{ch}$. In practice, the covariances used in NKL will have larger dimension since they will include angular correlations as well. However, this discussion will be clearer if we consider only the $n_{ch} \times n_{ch}$ case, and a mixing matrix plus template model can still be used to describe the case involving angular correlations. 

The signal model $\mathbf{C}_{S}$ includes the statistics of whatever components the analyst would like to recover from beneath the foregrounds. In this paper, we will take $\mathbf{C}_{S} = 
\mathbf{C}_{HI}$ unless otherwise noted. It should be noted that the freedom to choose $\mathbf{C}_{S} = 
\mathbf{C}_{HI}$ will provide NKL with an advantage over the other techniques mentioned in this paper. These techniques make no distinction between the HI and noise. The KL technique works by solving the generalized eigenvalue problem
\begin{equation} 
    \mathbf{C}_{FG}\mathbf{\Phi} = \mathbf{C}_{S} \mathbf{\Phi} \mathbf{\Lambda}.
\label{eq:generalized_eigen}
\end{equation}
In this formula, $\mathbf{\Phi}$ is a matrix of eigenvectors and $\mathbf{\Lambda}$ is a diagonal matrix of eigenvalues. It turns out that the eigenvectors $\mathbf{\Phi}$ obtained are a solution to the optimization problem \cite{gen_eig}
\begin{equation}
\begin{split}
\underset{\mathbf{\Phi}}{\max}  \; \mathbf{tr}(\mathbf{\Phi}^{T}\mathbf{C}_{FG} \mathbf{\Phi}) \\ 
\text{subject to} \; \mathbf{\Phi}^{T} \mathbf{C}_{S}\mathbf{\Phi} = \mathbf{I}.
\end{split}
\end{equation}
So, the KL transform finds modes that have as high a ratio of foreground to signal as possible. In particular, the eigenvalues $\lambda_{i}$ indicate the expected ratio of foreground to signal power at that particular mode. One then cleans the data by removing modes that are foreground dominated. In the language of templates and mixing matrices, we find that the template matrix is 
\begin{equation}
    \mathbf{S} = \mathbf{P}_{s} (\mathbf{X}-\overline{\mathbf{X}}).
\end{equation}
In this formula, $\mathbf{P}_{s}$ is a $n_{t} \times n_{ch}$ matrix whose rows are the foreground dominated eigenvectors. Moreover, let the symbol $\mathbf{P}$ to represent a $n_{ch} \times n_{ch}$ matrix whose rows are the eigenvectors generated by the generalized eigenvalue problem. The mixing matrix is then 
\begin{equation}
    \mathbf{A} = \mathbf{P}^{-1}_{s},
\end{equation}
where $\mathbf{P}^{-1}_{s}$ is a $n_{ch} \times n_{t}$ matrix containing only columns of $\mathbf{P}^{-1}$ that correspond to foreground dominated modes.

\subsection{Discussion of Techniques}
Before actually testing any techniques, it will be beneficial to discuss differences between the techniques considered in this paper. In particular, we will discuss these differences and try to provide some motivation for why certain techniques may provide better performance than others.



First, one should note that GNILC requires a model for the HI and noise, making it a ``non-blind" technique. This is different from GMCA, PCA and ICA, where the only free parameter provided by the user is the number of templates to use. 

Another aspect of these techniques to consider is locality. In the standard version of GMCA, the matrices $\mathbf{A}$ and $\mathbf{S}$ are meant to capture the foregrounds at all pixels and angular scales \cite{Carucci_2020}. PCA and ICA are typically conducted in a similar way, where one estimates the covariance $\mathbf{C}$ using the entire dataset. On the other hand, GNILC cleans the maps in a more fine-grained way. In particular, it divides the data into needlet coefficients, and then cleans the data one coefficient at a time. This allows for the foregrounds to be cleaned differently depending on the location and angular scale in question. 

One can imagine reasons why such a fine-grained treatment might provide advantages. For instance, the ratio of foreground to HI power is expected to vary with angular scale. In particular, large angular scales will suffer worse contamination than smaller ones. When working with unpolarized foregrounds, one would expect similar chromaticity at all lines of sight. However, when dealing with polarized foregrounds, we expect for the chromaticity to vary with line of sight. As such, one would expect for pixels close to the Galactic plane to require more templates to clean than ones far from the plane. Thus, it seems likely that a more fine-grained approach would work better when dealing with polarized foregrounds.

It is certainly possible to imagine changing PCA, ICA or GMCA to make them more local. For instance, one could implement a scale- and location- dependent version of PCA where the covariance matrix is estimated for some neighborhood around each needlet coefficient. In addition, a technique called L-GMCA has been proposed \cite{lgmca} in which one uses different mixing matrices for different regions of the map. For now though, we will consider only the global versions of these techniques. 

 So, we see reasons why a more localized approach would likely provide better results when cleaning foregrounds from maps. So far, GNILC is the most local approach that has been found in the literature. In \citet{Olivari_2015}, this approach was tested on low-z simulated maps without any polarization leakage included \cite{Olivari_2015}. This study found GNILC outperforming PCA at angular scales ($\ell>30$). However, GNILC provided worse performance than PCA at scales larger than that. Interestingly, \citet{Carucci_2020} found that GMCA also struggled at large angular scales $\ell < 50$, at least in the case when polarized foregrounds were included.

 In the next section, we introduce NKL. Similarly to GNILC, NKL is a non-blind technique which acts on needlet coefficients. However, while GNILC only considers statistics along the line of sight, NKL also considers angular correlations in needlet space.  
 




\section{Implementing NKL} \label{section:nkl}
In this section, we introduce the NKL technique for removing foregrounds from 21\,cm maps. We will begin this Section by providing background knowledge required for understanding how NKL works. In particular, the first Subsection introduces needlets. In the second, we describe the way in which needlet coefficients are partitioned before cleaning is performed. Then, we describe the ways in which we can generate an approximation of the foregrounds from the data. In the final subsection, we provide a list of steps for performing NKL. 



\subsection{The Needlet Transform}
The NKL process begins by performing a needlet transform on each frequency slice of the 3D maps. Needlets are wavelet-like functions that have a finite width in both $\ell$-space and pixel-space. These functions were first presented in \citet{Narcowich2006LocalizedTF}. For our purposes, we computed these coefficients using the \texttt{pys2let} software package, details of which can be found in \citet{Leistedt_2013}. Needlets are defined via 
\begin{equation}
    \psi_{jk}(\hat{n}) = \sqrt{\lambda_{jk}} \sum_{l} b\left(\frac{l}{B^{j}}\right)\sum_{m=-l}^{l} \overline{Y}_{lm}(\hat{n}) Y_{lm}(\xi_{jk}). 
\end{equation}
In this formula, $b\left(\frac{l}{B^{j}}\right)$ is a bandpass function that is non-zero for $B^{j-1} \leq l \leq B^{j+1}$. In this  paper, we choose $B=2$ and use $4 \leq j \leq 8$. The very lowest $\ell$ modes are described using a ``scaling function" \cite{Leistedt_2013}. The variable $\xi_{jk}$ refers to the line of sight at which the needlet is centered. In this paper, $\xi_{jk}$ refers to the location of pixel $k$ in a HEALPix map with $nside = 2^{j+1}$. The variable $\lambda_{jk}$ then refers to the solid angle of the HEALPix pixel in question. Needlet coefficients can then be obtained simply by computing the following integral
\begin{equation}
    \chi_{jk} = \int T(\hat{n}) \psi_{jk}(\hat{n}) d\Omega. 
\end{equation}
In this formula, $T(\hat{n})$ is the function for which the needlet coefficients are being computed. In this work, $T(\hat{n})$ will be the sky temperature. One can then reconstruct their temperature map using these coefficients as follows
\begin{equation}
    T(\hat{n}) = \sum_{jk} \chi_{jk} \psi_{jk}(\hat{n}). 
\end{equation}
\citet{Marinucci_2007} provides more detail on needlets and their use in cosmology.

\subsection{Partitioning of the Needlet Coefficients}
\label{subsection:chunking}
\begin{figure}
    \centering\includegraphics[scale=0.39]{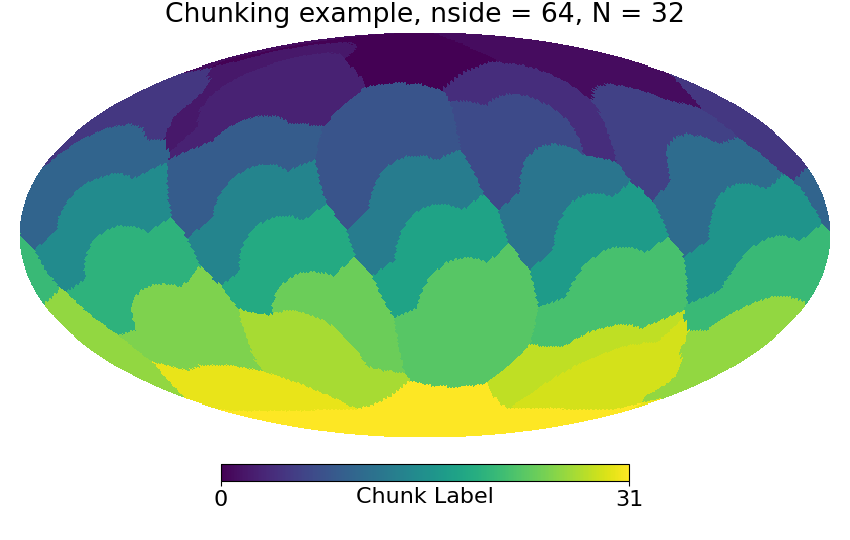}
    \caption{An example of dividing a needlet map into chunks.}
    \label{fig:chunking_example}
\end{figure}

An important part of NKL is the partitioning of the needlet coefficients. We refer to the groups of partitioned coefficients as ``chunks". In Fig.~\ref{fig:chunking_example}, we present an example of a chunked map. For now, let's suppose we are partitioning the coefficients for needlet scale $j$. Let's package the needlet coefficients at this scale of interest in a matrix $\boldsymbol{\chi}^{(j)}$. This matrix will have dimensions $n_{ch} \times n_{co}^{(j)}$, where $n_{ch}$ is the number of frequency channels and $n_{co}^{(j)}$ is the number of needlet coefficients per frequency channel at scale $j$.  

Next, let $\boldsymbol{\chi}^{(j)}_{i}$ be the row of $
\boldsymbol{\chi}^{(j)}$ corresponding to frequency channel $i$. The process of partitioning the row into $N$ chunks begins by selecting the coefficients located at columns $q = a*n_{co}^{(j)}/N$ for $a = 1,2,...,N$. Note here that $q$ is indexing pixels in the HEALpix map. Let's refer to these coefficients as ``anchors". Next, we assign all the other coefficients in the row to chunks according to which anchor point they have the smallest angular separation from. Adjacent neighborhoods then swap coefficients until all contain the same number of coefficients. 

We then use the same chunk assignments for all rows in $\boldsymbol{\chi}^{(j)}$. This results in a new ``chunked" matrix $\underline{
\boldsymbol{\chi}}^{(j)}$ which has dimensions $n_{ch}N \times n_{co}^{(j)}/N$. One row of $\underline{\boldsymbol{\chi}}^{(j)}$ will contain the coefficients of one chunk at one frequency channel. Column $q$ will contain the $q$\textsuperscript{th} element from each chunk at all frequency channels.

\subsection{Producing a Foreground Approximation} \label{subsection:fg_approx}
In order to perform a Karhunen-Loève transform, we need estimates for $\mathbf{C}_{FG}$ and $\mathbf{C}_{HI}$. The estimate of $\mathbf{C}_{HI}$ is provided as a prior chosen by the analyst. For the purposes of this paper, we estimate $\mathbf{C}_{HI}$ from maps simulated by CORA \cite{Shaw_2015}. One may also want to estimate $\mathbf{C}_{N}$ to include in the KL transformation or for later debiasing. In this paper, we estimate $\mathbf{C}_{N}$ from the model used to simulate the noise. In a real experiment, this noise covariance could be determined from the data. On the other hand, we seek to estimate $\mathbf{C}_{FG}$ from the data itself. This approach is motivated by two facts. First, as of now, polarized Galactic synchrotron radiation is not well enough measured or modeled to use an a priori covariance. In addition, a method that estimates the foregrounds directly from data will be more robust against errors in calibration and systematics that may cause the data to deviate from what would otherwise be an accurate a priori model. It should be noted that relatively small errors in the model of $\mathbf{C}_{FG}$ can cause significant problems, due to the large dynamic range between the foregrounds and HI. In this paper, two possible approaches to foreground estimation are presented, PCA and  DPSS Approximate lazY filtEriNg of foregroUnds (DAYENU) \cite{Ewall_Wice_2020}. In the DAYENU approach, one assumes that the foregrounds are confined within some region of delay space. This assumption still holds in our CRIME generated foreground maps, at least for lines of sight outside the Galactic plane. This assumption can be useful, since it means that the HI for larger delays will not be affected by the filter. This is different from PCA, where some signal loss outside the foreground-dominated region of delay space may occur. Moreover, unlike PCA, the DAYENU method treats each pixel (or needlet coefficient) uniquely. In PCA, one is using the same templates to approximate the foregrounds for all pixels/coefficients. The main weakness of DAYENU, however, is that it does not work when all delays are contaminated by foregrounds. 

\subsubsection{Estimating Foregrounds Using PCA}
This process begins by estimating the frequency-frequency covariance of the needlet coefficient maps. For this paper, we performed PCA foreground estimates one needlet scale at a time and without any chunking. PCA was not performed on chunked maps, as our scheme for selecting the number of components to remove struggled in this case. It's possible that applying PCA to each chunk individually would provide a better approximation, but we leave this to future testing. Once again, we are making the assumption that each line of sight is drawn independently and identically from some distribution. Since the needlet coefficients at a particular frequency channel have 0 mean, we can estimate their covariance as

\begin{equation}
    \hat{\mathbf{C}}^{(j)} = \frac{1}{n_{co}^{(j)}} \mathbf{\boldsymbol{\chi}}^{(j)} (\boldsymbol{\chi}^{(j)})^{T}.
\label{eq:cov}
\end{equation}
The only exception to this is for the ``scaling function" of the needlet transform, which contains the $\ell=0$ spherical harmonic mode. For the scaling function, one can just use equation~(\ref{eq:pca_cov}).
Due to their simpler frequency dependence, we will assume that the foregrounds are restricted to a subset of these modes. On the other hand, we will assume that the HI signal and noise ($\mathbf{s}^{(j)}$) will be more evenly spread out throughout all the modes. This naturally divides the eigenmodes into $n_{fg}$ modes which are dominated by the foregrounds and $n_{ch} - n_{fg}$ modes which are dominated by the signal plus noise. Let $\mathbf{P}$ be a matrix whose rows are eigenvectors of $\hat{\mathbf{C}}^{(j)}$. Moreover, let $\mathbf{P}_{f}$ be a $n_{fg} \times n_{ch}$ matrix containing the rows of $\mathbf{P}$ corresponding to the largest $n_{fg}$ modes of $\hat{\mathbf{C}}^{(j)}$. Next, let $(\mathbf{P}^{-1})_{f}$ be a $n_{ch} \times n_{fg}$ matrix whose columns are the columns of $\mathbf{P}^{-1}$ which act on the foreground dominated rows of $\mathbf{P}$. A foreground estimate can then be obtained via 

\begin{equation}
    \hat{\mathbf{f}}^{(j)} = (\mathbf{P}^{-1})_{f}\mathbf{P}_{f} \boldsymbol{\chi^{(j)}}. 
\end{equation}
The value for $n_{fg}$ can be estimated in different ways. In \citet{Zhang_2016}, a likelihood ratio test was used. In such a method, one essentially increases the number of parameters used until the resulting power spectrum begins to converge \cite{Zhang_2016}. Alternatively, one can use their models of the HI and noise, along with the Akaike Information Criterion (AIC), to estimate $n_{fg}$. This was the approach taken in \citet{Olivari_2015}. The AIC is given by 
\begin{equation}
    AIC = 2k - 2ln(L),
\end{equation}
where $k$ is the number of parameters in the model and $L$ is the maximum likelihood value of the model. The likelihood function used in this paper can be found in the appendix of  \citet{Olivari_2015}. Thus, by minimizing the AIC, one rewards goodness of fit (via the second term) while also discouraging overfitting (via the first term). More information on the use of AIC in a context such as this can be found in \citet{Olivari_2015}. In this paper, the number of modes chosen in any given situation will always be selected using AIC unless otherwise noted. 

\subsubsection{Estimating Foregrounds Using DAYENU}
Another possible approach to estimating the foregrounds is to use the DAYENU model from \citet{Ewall_Wice_2020}. DAYENU models the foreground covariance as simply as possible, using the fact that the foregrounds should be highly concentrated at delays close to $\tau = 0$. In particular, DAYENU will model the foreground covariance in delay space as 
\begin{equation} \label{eq:dayenu_tau}
    \tilde{C}^{\daleth}_{FG}(\tau,\tau') = \epsilon^{-1} \frac{1}{2\tau_{w}} \delta(\tau - \tau') \;\;\;\;\; |\tau| < \tau_{w}.
\end{equation}
In this equation, $2\tau_{w}$ is the assumed full width of the foregrounds in delay space and $\epsilon^{-1}$ is their assumed magnitude.

At a high level, our DAYENU-based scheme approximates the foregrounds at some line of sight in needlet space using the following steps:
\begin{enumerate}
    \item Estimate the width of the foregrounds in delay-space for the needlet coefficient of interest by comparing the delay transform to priors. The full width will correspond to $2\tau_{w}$.
    \item Use this width to generate the foreground covariance model $C^{\daleth}_{FG}(\nu,\nu')$. We would like to emphasize here that this foreground covariance is estimated one line of sight at a time. It is also not the same as the foreground covariance that will be eventually used in the KL step of NKL.
    \item Project the needlet coefficient onto the image of $C^{\daleth}_{FG}(\nu,\nu')$. This acts as a bandpass filter in delay-space, selecting the foreground-dominated region.
    \item Perform steps 1 through 3 at all coefficients in the needlet map. This will provide a foreground approximation $\hat{\mathbf{f}}_{k}^{(j)}$ from which one can estimate $\hat{\mathbf{C}}_{FG}^{(j)}$.
\end{enumerate}
This list has given a high level explanation of the DAYENU-based scheme. Let's now consider the details. 

First, recall the DAYENU covariance definition provided in equation~(\ref{eq:dayenu_tau}). In this formula, will use the value of $\tau_{w}$ acquired in the previous steps and will set $\epsilon=1$ for now. We will account for the magnitude of the foregrounds later on in this work.


In frequency space, the DAYENU foreground covariance will be given by 
\begin{equation} \label{eq:DAYENU}
     C^{\daleth}_{FG}(\nu,\nu') = \mathrm{sinc}(2\pi\tau_{w}(\nu-\nu')).
\end{equation}
When frequency samples are evenly spaced, $C^{\daleth}_{FG}(\nu,\nu')$ will be diagonalized by Discrete Prolate Spheroidal Sequences (DPSS) \cite{1978ATTTJ..57.1371S}. Suppose one is dealing with sequences of length $N_{d}$. The DPSS sequences $\{u^{\alpha}(N_{d},W) \mid \alpha = 1,2,3,\dots,N\}$ will form an orthonormal basis that maximizes energy in a region $[-W,W]$ of the DFT domain. We call this ``spectral concentration" and denote it as $s_{c}$. More precisely, the spectral concentration for some sequence is
\begin{equation}
   s_{c} =  \frac{\int_{-W}^{W} |U(x)|^{2} dx }{\int_{-1/2}^{1/2} |U(x)|^{2} dx}. 
\label{eq:spec_conc}
\end{equation}
In this formula, $x$ is the variable of the DFT domain, which in our context is delay. The variable $U$ represents the DFT of $u$, centered at $(N_{d}-1)/2$. $u^{1}$ is defined as a unit norm sequence that maximizes $s_{c}$. Then, $u^{2}$ is created by finding a unit norm sequence that maximizes $s_{c}$ while being orthogonal to $u^{1}$. In general, $u^{n}$ is created by finding a unit norm sequence that maximizes $s_{c}$ while being orthogonal to $u^{i}$ for $i = 1,..,n-1$. 

In the case of DAYENU, the DPSS sequences will maximize energy at delays $-\tau_{w} < \tau < \tau_{w}$. 

The DAYENU model can then be used to estimate the foregrounds for each needlet coefficient in the following way. For the following steps, let $\mathbf{\xi}_{k}^{(j)}$ be a $n_{ch} \times 1$ vector containing needlet coefficients for all channels at scale j and pixel k. The procedure to estimate the foregrounds at $\chi_{k}^{(j)}$ is as follows.

\begin{enumerate}
\item Use priors on the HI and noise to generate mock needlet coefficient maps. These mock maps will be denoted by $\mathbf{h}^{m,(j)}$ and $\mathbf{n}^{m,(j)}$ for the HI and noise respectively. Moreover, let $\mathbf{y}^{m,(j)} = \mathbf{h}^{m,(j)} + \mathbf{n}^{m,(j)}$.
\item Randomly select $N_{p}$ lines of sight from these mock coefficient maps. Then, apply two Blackman-Harris windows and an FFT to bring them into the delay domain. Moreover, the data should be shifted to place $\tau = 0$ at the center sample. We will denote such a transform with a tilde, i.e. 
\begin{equation}
    \tilde{a} = SHIFT[FFT[aW^{2}]],
\end{equation}
where $W$ denotes a four term Blackman-Harris window of appropriate length. Let $\mathcal{D}$ be a set containing the randomly chosen pixels. 
\item For each $i \in \mathcal{D}$, perform a curve fit on $abs(\tilde{\mathbf{y}}_{i}^{m,(j)})$. This curve fit is meant to give one an estimate of the typical level of $abs(\tilde{\mathbf{y}}_{i}^{(j)})$ at each delay. It was found that using a third order polynomial in log-space worked quite well. The curve produced by this fitting process at pixel $i$ will be denoted $r_{i}^{m,(j)}$.
\item Average the fits obtained at each pixel to produce a typical HI plus noise curve in delay space:
\begin{equation}
    \overline{r}^{(j)} = \frac{1}{N_{p}} \sum_{i \in \mathcal{D}} r_{i}^{m,(j)}.
\end{equation}
In this formula, $N_{p}$ is the number of pixels chosen and $\mathcal{D}$ is the set of pixels used in the approximation.  
\item Perform a delay transform on the needlet coefficient of interest, giving $\tilde{\chi}_{k}^{(j)}$. Next, compare $abs(\tilde{\chi}_{k}^{(j)})$ with $\overline{r}^{(j)}$. The delay at which the level of $abs(\tilde{\chi}_{k}^{(j)})$ becomes comparable to that $\overline{r}^{(j)}$ will correspond to the $\tau_{w}$ used in DAYENU. For this paper, we select the delay at which $abs(\tilde{\chi}_{k}^{(j)})$ is within a factor of 2 of $\overline{r}^{j}$. 

\item Use the $\tau_{w}$ obtained in the previous step to generate a DAYENU covariance as in equation ~(\ref{eq:DAYENU}). This DAYENU matrix will be denoted via $\mathbf{C}^{\daleth}$. 

\item Compute the SVD of $\mathbf{C}^{\daleth}$, giving $\mathbf{C}^{\daleth} = \mathbf{U}\mathbf{S}\mathbf{V}^{\dagger}$. This SVD will be used to project onto the image of $\mathbf{C}^{\daleth}$. Let $\overline{\mathbf{U}}$ be a matrix containing columns of $\mathbf{U}$ for which the singular value is above some threshold. This threshold is determined based on the expected magnitude of the foregrounds. Since the foregrounds can be 5 orders of magnitude brighter than the signal, it is advisable to choose the cutoff to be at or below $10^{-10}s_{max}$, where $s_{max}$ is the largest singular value of $\mathbf{C}^{\daleth}$. In this particular work, we use the same threshold for all angular scales and all needlet locations. The foregrounds are then approximated via
\begin{equation}
    \hat{f}_{k}^{(j)} = \overline{\mathbf{U}} \overline{\mathbf{U}}^{T} \chi_{k}^{(j)}.
\end{equation}

\end{enumerate}

\subsection{The NKL algorithm}
In the literature, one finds standard foreground removal methods such as PCA, ICA and GMCA being used in a way where one pair of $\mathbf{A}$ and $\mathbf{S}$ is generated for the entire data set. One exception to this in the literature is GNILC, which performs the cleaning in a more fine-grained way, sending the data to the needlet domain and then cleaning each needlet coefficient individually. However, this method only takes advantage of frequency-frequency correlations in the data. Similarly to GNILC, NKL performs its cleaning in the needlet domain. The key difference, however, is that NKL also takes advantage of angular correlations in the needlet coefficients. Moreover, instead of using an ILC step, as in GNILC, NKL uses a KL transform. Below are step by step instructions for performing an NKL cleaning on some dataset.

\begin{enumerate}
    \item Generate an approximation of the foregrounds at the needlet scale of interest. In this paper, foreground approximations are generated using either DAYENU of PCA. This approximation will be denoted via $\hat{\mathbf{f}}^{(j)}$. 
    \item Divide the needlet coefficients at the scale of interest into N ``chunks" of equal size. These chunks must be adequately large to provide accurate covariance estimates. The appropriate chunk size will depend both on the severity of the foregrounds and on the needlet scale in question. This gives $\underline{\chi}^{(j)}$, which was defined in Section~\ref{subsection:chunking},
    \item Divide the foreground approximation into chunks, providing $\underline{\hat{\mathbf{f}}^{(j)}}$. Use $\underline{\hat{\mathbf{f}}^{(j)}}$ to generate an estimate of the foreground covariance at the scale of interest. This covariance can be estimated in the manner shown in equation~(\ref{eq:pca_cov}). This covariance will be denoted $\hat{\mathbf{C}}^{(j)}_{FG}$ and will be a square matrix of size $n_{ch}N$.




\item Use priors on the signal to create an estimate of $\mathbf{C}_{S}^{(j)}$. This matrix will have the same shape as the foreground covariance matrix described in the previous step.
\item Use the covariance matrices acquired in the previous two steps to perform a Karhunen-Loève transform on the needlet scale of interest. The steps for this process are described in Section~\ref{subsection:fgremoval}.  

\end{enumerate}

\subsection{Deviations From the Ideal Case} \label{subsection:negbias}
The NKL technique is based on the KL transform, which is described in equation~(\ref{eq:generalized_eigen}). There are two deviations from this ideal case that we will consider here. The first is inaccuracies in our estimates of $\mathbf{C}_{S}$ and $\mathbf{C}_{FG}$. The signal covariance, $\mathbf{C}_{S}$, is derived from priors and thus may be sensitive to the model chosen. We tested this in one case by using priors generated by CORA on maps generated using CRIME. We do not present the results in this paper, but we found that there was a negligible difference in performance.  

On the other hand, $\mathbf{C}_{FG}$ is estimated from the data. This foreground approximation will inevitably be contaminated by some signal and noise. More precisely, for scale $j$, we have 
\begin{equation}
    \hat{\underline{\mathbf{f}}}^{(j)} = \mathbf{\underline{f}}^{(j)} + \underline{\mathbf{h}}'^{(j)} + \underline{\mathbf{n}}'^{(j)}.
\end{equation}
In this formula, $\underline{\mathbf{h}}'^{(j)}$and $\underline{\mathbf{n}}'^{(j)}$ represent the  HI and noise present in the foreground approximation, respectively. We include underlines on these variables to emphasize that these maps have been partitioned as described in Section~\ref{subsection:chunking}. In this section, we will consider the effect that $\underline{\mathbf{h}}'^{(j)}$ and $\underline{\mathbf{n}}'^{(j)}$ have on the statistics of the cleaned coefficients. 

Let's begin by considering the generalized eigenvalue problem of equation~(\ref{eq:generalized_eigen}). For now, let's assume perfect knowledge of the foregrounds, noise and signal. Moreover, let us consider only one eigenmode, $\phi_{i}$, which has a corresponding eigenvalue, $\lambda_{i}$. Lastly, let's consider an analysis conducted only at one needlet scale, $j$. We will drop any $j$ subscripts since the analysis is all done at the same scale. This becomes the generalized eigenvalue problem   
\begin{equation} \label{eq:eigenvalsolve}
    \mathbf{C}_{FG} \mathbf{\phi}_{i} = \lambda_{i} \mathbf{C}_{S} \mathbf{\phi}_{i}.
\end{equation}
However, when performing NKL, we will use $\hat{\mathbf{C}}_{FG}$. We use the hat symbol to denote covariance estimates derived from the data. This gives a new eigenvalue problem 
\begin{equation} \label{pert_eig} 
    \hat{\mathbf{C}}_{FG} \mathbf{\phi}_{i}' = \lambda_{i}' \mathbf{C}_{S} \mathbf{\phi}_{i}',
\end{equation} 
where apostrophes represent perturbed versions of the variables seen in equation~(\ref{eq:eigenvalsolve}).  This estimate of the foreground covariance will vary from the true version as 

\begin{equation}\label{eq:fg_errors}
    \hat{\mathbf{C}}_{FG} = \mathbf{C}_{FG} + \mathbf{\Delta} _{FG} + \mathbf{C}_{h'h'} + 
    \mathbf{C}_{fh'} + \mathbf{C}_{n'h'} +  \mathbf{C}_{n'n'} +  \mathbf{C}_{fn'}.
\end{equation}
In this formula, $\mathbf{\Delta}_{FG}$ represents foreground errors that would be present if $\hat{\mathbf{C}}_{FG}$ were acquired from the true foregrounds. The $\mathbf{C}_{fh'}$ term is given by
\begin{equation}
    \mathbf{C} = \frac{1}{N_{p}} \sum_{q} (h_{q}' f_{q}^{T} + f_{q}h_{q}'^{T}),
\end{equation}
In this formula, $f_{q}$ is the q\textsuperscript{th} column of $\underline{\mathbf{f}} - \underline{\overline{\mathbf{f}}}$ and $h_{q}'$ is defined as $\underline{\mathbf{h}}' - \underline{\overline{\mathbf{h}}}'$. Note that the overline here has the same meaning as in equation~(\ref{eq:pca_cov}). The variable $N_{p} = n_{co}^{(j)}/N$ is the number of columns in $\underline{\mathbf{f}}'$ and $\underline{\mathbf{s}}'$. This is the same as the number of needlet coefficients per chunk per frequency channel. Note that the other terms in equation~(\ref{eq:fg_errors}), e.g. $\mathbf{C}_{n'h'}$, are defined in a similar way. 

Each of these terms ought to be a source of error in our resulting estimate of the HI signal. However, the most concerning of these is $\mathbf{C}_{fh'}$, which will lead to a negative bias in our estimate of the HI signal. In principle, $\mathbf{C}_{n'h'}$ should also lead to correlations between the noise and HI after cleaning has been done. However, we did not observe this leading to any noticeable bias in the resulting power spectra. Thus, we will focus on the contribution of $\mathbf{C}_{fh'}$. 

Let $\mathcal{F}_{i}'= \phi_{i}'^{T}f_{p}$ and $\mathcal{H}_{i}' = \phi_{i}'^{T} h_{p}$. We will find that $\mathbf{C}_{fh'}$ will lead to a non-zero correlation between $\mathcal{H}_{i}'$ and $\mathcal{F}_{i}'$. It turns out that this bias will be more severe for modes with larger eigenvalues $\lambda_{i}$. This comes into the picture through the variable 
\begin{equation}
\alpha_{ik} = E[\overline{\mathcal{H}}_{i}'\mathcal{H}_{k}'] \approx \begin{cases} 
      0 & \lambda_{i} < 1 \\
      \delta_{ik} & \lambda_{i} > 1,  
   \end{cases}
\end{equation}
where $E[]$ refers to an expectation value. It will be the case that 
\begin{equation} \label{eq:fisi}
      E[\mathcal{F}_{i}'\mathcal{H}_{k}'] \approx \delta_{ik} \frac{1}{N_{p}} \Sigma_{j \neq i} \frac{ \lambda_{j}\alpha_{ii} + \alpha_{jj}\lambda_{i}}{\lambda_{i}-\lambda_{j}}.
\end{equation}
In this equation, $\delta_{ik}$ is a Kronecker delta. A derivation of equation~(\ref{eq:fisi}) can be found in the Appendix. We emphasize that this is a rough estimate, but it does provide some insight into the statistics of NKL. There are four important features about equation~(\ref{eq:fisi}) that we would like to emphasize. 
\begin{enumerate}
    \item Signal-dominated modes ($\lambda_{i} < 1$) will be negatively biased, as they only pick up contributions from terms where $\lambda_{j} > \lambda_{i}$. 
    \item The $\lambda_{i}-\lambda_{j}$ in the denominator ensures that the bias will be small for modes $\lambda_{i} << 1$. However, the bias will be worse for modes with $\lambda_{i}$ closer to 1. 
    \item This bias scales like $\frac{1}{N_{p}}$, meaning that it will become less severe when larger chunks are used in the analysis. 
    \item The bias includes no dependence on the number of frequency channels used in the analysis. This is different from ILC methods in which the bias becomes more severe as more frequency channels are included.
\end{enumerate}
A test of this equation will be presented in Section~\ref{subsection:highz_test}.


\subsection{MGNILC, a corollary}
One can also use the foregrounds approximation methods of Section~\ref{subsection:fg_approx} to modify the GNILC technique described in \citet{Olivari_2015}. In GNILC, one generates an estimate for the HI signal plus noise by applying a filter to the data:

\begin{equation} \label{eq:gnilc}
    \hat{s} = \hat{\mathbf{S}}(\hat{\mathbf{S}}^{T} \hat{\mathbf{R}}^{-1} \hat{\mathbf{S}})^{-1} \hat{\mathbf{S}}^{T} \hat{\mathbf{R}}^{-1} x.
 \end{equation}
In this formula, $x$ is the data vector (all frequency channels for a particular needlet coefficient), $\hat{s}$ is an estimate of the HI plus noise for a particular line of sight, $\hat{\mathbf{R}}$ is an estimate of the covariance of the data and $\hat{\mathbf{S}}$ is given by 
\begin{equation}
    \hat{\mathbf{S}} = \hat{\mathbf{R}}_{HI+N}^{\frac{1}{2}} \mathbf{U}_{s}.
\end{equation}
Here, we have that $\hat{\mathbf{R}}_{HI+N}$ is a prior of signal plus noise covariance, and $\mathbf{U}_{s}$ is a matrix containing the subset of eigenvectors of $\hat{\mathbf{R}}$ that are dominated by the signal plus noise. In this formula, raising a matrix to the $1/2$ power refers to taking the Hermitian square root. When using an ILC technique such as GNILC, one must consider errors in $\hat{\mathbf{R}}$. In particular, $\hat{\mathbf{R}}$ will have some error due to the finite sample size used in the estimation:

\begin{equation}
    \hat{\mathbf{R}} = \mathbf{R} + \mathbf{\Delta}.
\end{equation}
In this formula, $R$ is the true covariance of the data and $\mathbf{\Delta}$ is an error term given by 

\begin{equation}
    \mathbf{\Delta} = (\hat{\mathbf{R}}_{HI+N} - \mathbf{R}_{HI+N}) + (\hat{\mathbf{R}}_{FG} - \mathbf{R}_{FG}) + \tilde{\mathbf{C}}.
\end{equation}
In this formula, the first two terms correspond to errors in the foregrounds and HI + n respectively. The third term corresponds to a spurious correlation that appears between the foregrounds and signal plus noise due to the finite sample size used. This is given by the equation
\begin{equation}
    \tilde{\mathbf{C}} = \frac{1}{N_{p}} \sum_{q} (s_{q}f_{q}^{T} + f_{q}s_{q}^{T}).
\end{equation}
In this formula, $N_{p}$ is the number of needlet coefficients in the neighborhood being considered and $q$ is a variable that indexes lines of sight in the needlet coefficient neighborhood. We also introduce the variable $s_{q} = h_{q} + n_{q}$. 

Similarly to what was seen in NKL, $\hat{\mathbf{C}}$ causes residual foregrounds to have a negative correlation with the HI plus noise. This leads to the negative ``ILC bias" described in \citet{delabrouille_2009}. This bias scales like $\frac{n_{ch}}{N_{p}}$. Such a bias is troublesome at larger scales, for which $N_{p}$ is restricted by the number of needlet coefficients available.

To mitigate this problem, we propose Modified GNILC (MGNILC). In this approach, we use a modified covariance estimate

\begin{equation}
    \hat{\mathbf{R}}_{m} = \hat{\mathbf{R}}_{FG} + \hat{\mathbf{R}}_{HI+N}.
\end{equation}
In this formula, $\hat{\mathbf{R}}_{FG}$ is an estimate of the foreground covariance in the needlet neighborhood of interest using the approach described in Section~\ref{subsection:fg_approx}. In this case, our error term will look like 
\begin{equation}
    \hat{\mathbf{C}}_{m} = \frac{1}{N_{p}}  (s_{q}'f_{q}^{T} + f_{q}s_{q}'^{T}).
\end{equation}
In this formula,  $s_{q}'$ is the HI left over after performing the foreground approximation. Whether using PCA or DAYENU for the foreground approximation, we should find that $s_{q}'$ is essentially a version of $s_{q}$ that has been low-pass filtered in delay space. This reduces the ILC bias in two ways. For one, the fact that the HI plus noise has been low-pass filtered prevents the negative bias from affecting larger delays. Moreover, it also prevents any spurious correlations between high delay components of $s$ with the low delay components of $f$, providing some additional mitigation of the ILC bias. However, the ILC bias problem will not be fixed completely as there will of course still be correlations between $f_{q}$ and the low delay content of $s_{q}$.  

\section{Simulated Maps For Testing}\label{sec:sim_maps}
This section provides a description of the simulated maps used for testing the various foreground removal techniques described earlier in this paper. 

\subsection{Cosmological Signal}
We generated simulated HI maps using the CORA package \footnote{\url{https://github.com/radiocosmology/cora}}. The CORA software assumes that the 21\,cm signal is Gaussian and isotropic. For computational convenience, the covariance is estimated using the flat sky approximation. One then simulates the 21\,cm fluctuations using equation C5 from \citet{Shaw_2015}. The mean HI temperature is then supplied using equation C4 from that same paper. The CORA software uses cosmological parameters from Planck 2018 \cite{2020A&A...641A...6P}. Moreover, we used the default CORA setting, which is to estimate $\Omega_{HI}$ using the model given in \citet{Crighton_2015}. The CORA software package also incorporates redshift space distortions, assuming a constant HI bias of $b(z) = 1$ by default.

\subsection{Foregrounds}
In this paper, we consider only foregrounds from Galactic synchrotron radiation. For the synchrotron radiation, both unpolarized and polarized contributions are included. For the polarized foregrounds, 1 percent leakage of both the Q and U components was assumed. This value was chosen to be consistent with a typical level of polarization leakage for HI intensity mapping telescopes. These foregrounds are simulated using the CRIME software package \cite{Alonso_2014}. CRIME models unpolarized Galactic synchrotron emission via 
\begin{equation}
    T(\hat{n},\nu) = T_{\mathrm{Haslam}}(\hat{n})(\frac{\nu}{\nu_{H}})^{\beta(\hat{n})} + \delta T.
\end{equation}
In this formula, $T_{\mathrm{Haslam}}$ is the Haslam map temperature, $\beta$ is a spectral index generated from the Planck sky model \cite{refId0} and $\nu_{H}$ is 408\, {\rm MHz}, the frequency of the Haslam map. The $\delta T$ term accounts for angular scales $l \gtrsim 200$, which are smaller than the Haslam map's resolution. This term is generated assuming that the foregrounds are Gaussian and isotropic at these scales. The power spectrum assumed can be found in  \citet{Alonso_2014}. 

The polarized synchrotron emission is modelled using the measured Faraday depth through the entire Milky Way presented in \citet{Oppermann_2012}. CRIME then assumes, among other things, that the number of emitting regions follows a Gaussian distribution as a function of Faraday depth.  

CORA also can be used to model the polarized synchrotron radiation. The model used by CORA is essentially the same as that used by CRIME. The main theoretical difference is that CORA assumes that emitting regions at the same Faraday depth for a particular line of sight are independent. This leads to a slightly different dependence on the emission at Faraday depth $\psi$ for a particular line of sight. The Faraday depth coherence lengths are also slightly different, being $0.5\,{\rm rad/m^{2}}$ for CRIME and $1\, {\rm rad/m^{2}}$ for CORA. The value for CRIME was chosen to provide results consistent with the Hammurabi simulation package \cite{hammurabi}, while CORA's value was chosen to be consistent with observations at 1.4\,GHz \cite{Wolleben_2006}. In practice, we found that the foregrounds provided by CRIME appeared to be more chromatic than those from CORA. Thus, to be conservative, we used CRIME to simulate the foregrounds.   

\subsection{Instrumental Effects and Noise} \label{subsection:insteff}
In this paper, tests of various foreground removal techniques are performed assuming an instrument similar to the MeerKAT array operating in single dish mode.  The parameters of this telescope are described in Table~\ref{table:telescope}. One may notice that the observing time of $40000$\,{\rm hrs} is quite large. This value is not particularly realistic, but was chosen to ensure that the noise would not overpower the HI signal. In future telescopes such as PUMA, adequate noise levels can be achieved with larger numbers of antennas and less integration time. The value $f_{sky} = 1$ was chosen since we are imagining an instrument that observes the whole sky. Any masking is then applied after observations have taken place. 

This hypothetical instrument was chosen for realism and simplicity. Since it consists of an array of dishes operating in single dish mode, we can model the beams by convolving with a Gaussian. This is much easier than trying to derive maps from simulated visibilities. Moreover, the relatively low angular resolution of the instrument ($\approx 1^{\circ}$) ensures that large values of $\ell$ will not be required. This relaxes computational requirements when working with needlets. We approximate beam effects in a way that is identical to \citet{Carucci_2020}. In particular, we convolve the simulated maps with a frequency-dependent Gaussian beam. We then reconvolve all maps to give them the same angular resolution.

\begin{table} 
\centering
\begin{tabular}{cc}
\hline
\hline
\textbf{Parameter}&
\textbf{Value} \\
\hline

D & 13.5\,m \\
$T_{inst}$ & 20\,K \\
$f_{sky}$ & 1\\
$t_{obs}$ & 40000\,hrs \\
$N_{dishes}$ & 64 \\
$n_{ch}$ & 256 \\ 
$[\nu_{min},\nu_{max}]$ & $[980,1080\,{\rm MHz}]$ and $[400,500\,{\rm MHz}]$.
\\
$\Delta \nu$ & $0.390625\,{\rm MHz}$ \\
\hline
\hline
\end{tabular}
\caption{Parameters describing the hypothetical telescope used in this paper.}
\label{table:telescope}
\end{table}
We added noise to the simulated maps in the same way as Carucci. In particular, the noise per pixel follows a Gaussian distribution with a standard deviation given by 

\begin{equation} \label{eq:sign}
    \sigma_{N}(\nu) = T_{\mathrm{sys}}(\nu) \sqrt{\frac{4 \pi f_{\mathrm{sky}}}{\Delta \nu t_{\mathrm{obs}} N_{\mathrm{dishes}} \Omega_{\mathrm{beam}}}}.
\end{equation}
In this formula, $T_{sys}$ is the system temperature, $f_{sky}$ is the fraction of the sky observed in the survey, $\Delta \nu$ is the frequency resolution, $t_{obs}$ is observing time and $\Omega_{beam}$ is the beam solid angle. Similarly, we estimate the system temperature in the same way as Carucci, assuming
\begin{equation}
    T_{sys}(\nu) = T_{instr}[\mathrm{K}] + 66\big(\frac{\nu}{300 [\mathrm{MHz}]}\big)^{2.55}. 
\end{equation}

Once noise and Gaussian beam effects are accounted for, the maps used in testing can be described schematically via the equation 

\begin{equation}
    \mathbf{X} = (B_{low}-B)*(B*(\mathbf{f} + \mathbf{h}) + \mathbf{n}).
    \label{eq:schematic}
\end{equation}
In this schematic, $B*$ represents convolution with the beam and $B_{low}$ is the beam at lowest frequency channel in the band.

\section{Tests of Methods} \label{section:tests}
In this section, we test the performance of various foreground removal methods and compare the results. In particular, we test GMCA, PCA, ICA, GNILC, and NKL. It should be noted that the severity of polarized foregrounds is expected to vary as a function of redshift. In particular, the chromaticity of polarized foregrounds will become more severe at lower frequencies due to Faraday rotation. Therefore, we conduct tests in two different redshift regimes corresponding to low frequency and high frequency bands of MeerKAT.  

\subsection{Additional Map Preparation Steps}
After performing the steps described in Section \ref{subsection:insteff}, we took  some additional steps to aid the cleaning process. These were done both to exclude highly contaminated pixels and to lower computation times. 

\begin{figure}
    \centering
    \includegraphics[scale=0.3]{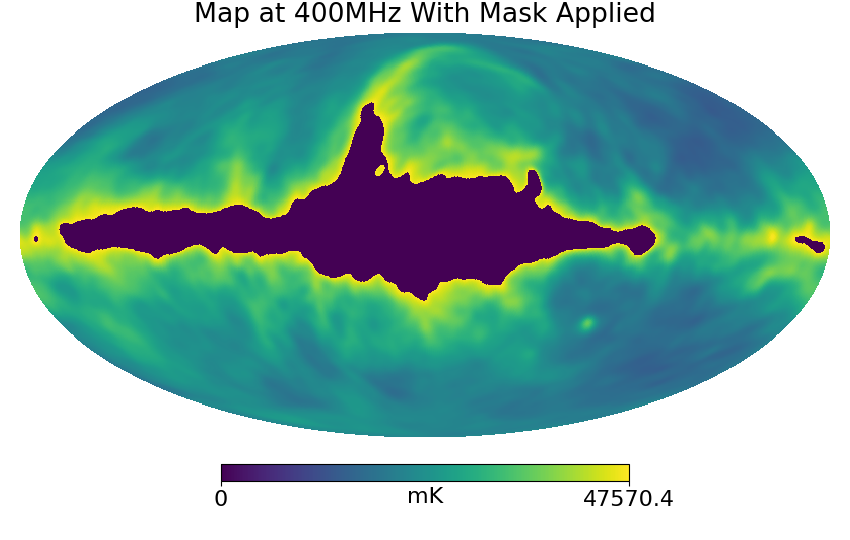}
    \caption{An illustration of the mask used for tests of the foreground removal methods in the high redshift case. The mask has been applied to this map, which includes simulated foregrounds, signal and noise.}
    \label{fig:mask}
\end{figure}

\begin{figure}
    \centering
    \includegraphics[scale=0.5]{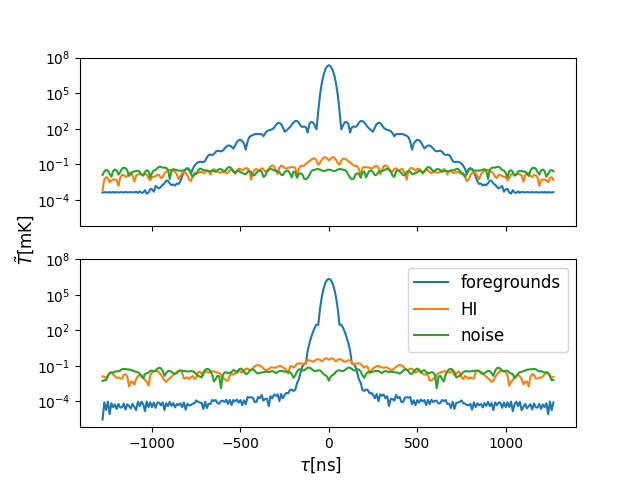}
    \caption{The top plot shows delay spectra at the brightest pixel in the unmasked map in the higher redshift case. The bottom plot shows delay spectra at the brightest pixel remaining after the mask has been applied.}
    \label{fig:delay_comparison}
\end{figure}

As expected, polarized foreground chromaticity is quite severe in the Galactic plane. We applied a mask to the data to exclude the brightest 15 percent of pixels. Such masks were created for both the low-z and high-z test cases. The mask created for the high-z test case is shown in Fig.~\ref{fig:mask}. The chromaticity problem is illustrated in Fig.~\ref{fig:delay_comparison}. The top plot in the figure shows delay spectra for the brightest line of sight in the high-z test case. Note here that the foregrounds dominate for almost all delays. The bottom plot shows delay spectra at the brightest unmasked line of sight. In this case, we find that the foregrounds occupy a relatively small region of delay space, allowing for more effective cleaning. 

It was noted that the Gaussian half-power beam-widths in both the low-z and high-z test cases are greater than $ 1^{\circ}$. As such, we truncated maps in both test cases at $\ell=255$ for computational simplicity. Moreover, we performed needlet transformations using $B=2$ with scales running from $j=4$ up to $j=8$. We provide details of how the data is represented in the needlet domain in Table~\ref{table:needlets}.

It is worth discussing here the impacts that $B$ and the range of $j$ can have on one's analysis. Increasing the value of $B$ will decrease the spread of the needlet coefficients in real space while increasing their spread in harmonic space. So, choosing a value of $B$ greater than 2 would in principle allow for finer chunking of the maps but would also lead to coarser partitioning in $\ell$-space.

Next, let's consider the range of $j$ chosen. Since the maps were truncated at $\ell =255$, scales with $j>8$ would contain no information and thus were not included. The minimum scale value $j_{min}=4$ was chosen to ensure that the scaling function would have enough statistics to be cleanable by NKL.

The value of $B$ and range of $j$ chosen for this paper do work. However, other values were not tested, so it is unknown whether these are the optimal choices one could make.

\begin{table}
\centering
\begin{tabular}{ccc}
\hline
\hline
\textbf{Scale}&
\textbf{l range}&
\textbf{nside} \\
\hline

Scaling Function & $[0,16]$ & 16 \\
$j=4$ & $[8,32]$ & 32 \\
$j=5$ & $[16,64]$ & 64\\
$j=6$ & $[32,128]$ & 128 \\
$j=7$ & $[64,256]$ & 256 \\ 
$j=8$ & $[128,512]$ & 512 \\
\hline
\hline
\end{tabular}

\caption{A summary of the needlet coefficients in these tests.  We use the same scales, $l$ ranges and nside values for both the low-z and higher-z tests.}
 \label{table:needlets}
\end{table}

\subsection{Implementation of Cleaning Techniques}
When cleaning with blind methods like PCA, ICA and GMCA, we use priors on the HI signal and noise to select the appropriate number of components to be removed. In particular, we generated additional sets of HI maps using CORA and noise maps using equation~(\ref{eq:sign}). We then applied the beam convolution described in Section~\ref{subsection:insteff} and  produced covariance estimates from these processed maps using equation~(\ref{eq:cov}). 
In this paper, we chose to use $N_{maps} = 10$ mock maps for estimating the covariances. We then used our models of the HI and noise covariances to estimate the appropriate number of modes to remove using the AIC prescription described in Olivari \cite{Olivari_2015}. 

For the cases of GNILC, MGNILC and NKL, we used the same sets of mock HI and noise maps as above to estimate the needlet coefficient covariances.

For GNILC, we selected the appropriate number of coefficients using the Olivari AIC prescription. We also chose to use windows with a size of at least $10^{5}$ coefficients. This seemed to be the window size required to minimize the ILC bias described in \citet{delabrouille_2009}. There were of course exceptions to this rule for scales up to $j=6$, since needlet maps at lower $j$ did not have enough coefficients. These windows were generated simply by choosing the closest $N$ coefficients to the coefficient of interest.

\subsection{Power Spectrum Estimation}
In order to estimate the power spectrum, we used the method described in \citet{Liu_2016}. In this approach, one computes Bessel-Spherical harmonic modes of the sky
\begin{equation}
    T_{\ell m}(k) = \sqrt{\frac{2}{\pi}} \int d\Omega dr r^{2} j_{\ell}(kr) Y_{\ell m}^{*}(\hat{n}) \phi(\mathbf{r}) T(\mathbf{r}),
\end{equation}
where $\phi(\mathbf{r})$ is a window function representing the survey volume. One then computes a ``spherical harmonic power spectrum" 
\begin{equation}
    S_{\ell}(k) = 2 \pi^{2} \left[ \int d^{3} r \phi(\mathbf{r})^{2} j_{\ell}(kr)^{2}\right]^{-1}  \frac{\Sigma_{m} |T_{\ell m}(k)|^{2}}{2\ell + 1}.
\end{equation}
This spherical harmonic power spectrum is a spherical analogue to the commonly seen cylindrical power spectrum. It is also a useful tool in that it allows for different $\ell$-modes  to be checked individually. Then, in the case of a  translation-invariant sky, one can form the power spectrum estimator 
\begin{equation} \label{eq:power}
    \hat{P}(k) = \Sigma_{\ell} w_{\ell} S_{\ell}(k). 
\end{equation}
In this formula, the $w_{\ell}$ are weights which depend on the survey volume used. These weights account for the fact that sensitivity to a certain $k$-mode may vary with angular scale $\ell$. In this case of a translation-invariant sky, $\langle \hat{P}(k) \rangle = P(k)$. It should be mentioned that the maps presented in this paper do not strictly satisfy translation invariance. Foregrounds vary significantly with line of sight and the HI is subject to redshift-space distortions and cosmic evolution. Even the noise breaks translation invariance, as its amplitude varies with frequency. Despite this, we still present $\hat{P}(k)$ as a ``power spectrum", since it is provides us with a weighted average of fluctuations at length scale $k$, even when translation invariance is broken.

\subsection{Test at Higher Redshifts} \label{subsection:highz_test}
For this test, we used the hypothetical instrument described in Table~\ref{table:telescope} and assumed it to have 256 evenly spaced frequency channels running from 400\,{\rm MHz} to 500\,{\rm MHz}.  

\begin{figure}
    \centering
    \includegraphics[scale=0.46]{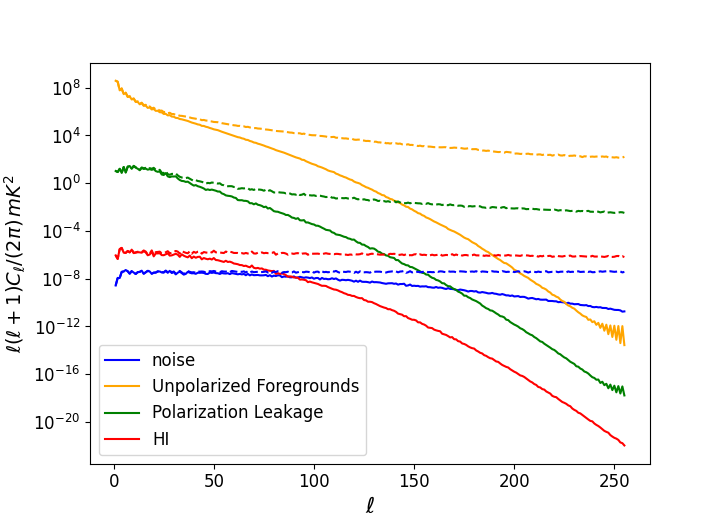}
    \caption{Angular power spectra estimated from unmasked maps using \texttt{healpy} at $450\,{\rm MHz}$. Solid lines correspond to maps which have undergone beam convolution. Dashed lines correspond to maps that have not undergone any convolution steps.}
    \label{fig:cl_highz}
\end{figure}

In Fig.~\ref{fig:cl_highz}, we provide angular power spectra for the $450\,{\rm MHz}$ frequency channel used in this test. The maps used to produce these power spectra have not been masked. We chose to use unmasked maps here in order to avoid edge effects at the mask boundaries. In this case, the beam convolution causes significant loss of signal above $\ell \approx 50$. Due to this, any power spectra produced by our instrument will be missing significant amounts of small length scale information. For this reason, we will evaluate the effectiveness of the cleaning techniques based on how well they recover the power spectra of convolved maps. It should also be noted that the noise suffers less loss at $\ell>50$ than the other components. This is due to the fact that the foreground and HI have been convolved with a beam twice while the noise has only been convolved once. This causes the noise to be the dominant component of the maps at those smaller angular scales. 

For the case of NKL, the chunk size and desired SNR of the KL cleaning vary with needlet scale. We summarize this in Table \ref{table:nkl_params_highz}. The values shown in this table were chosen through trial and error. Ideally, one would like to clean with as many chunks as possible, since this would provide the best statistics. However, we are limited by two factors. For one, making the chunks too small leads to an inaccurate $\hat{C}_{FG}$ and incomplete cleaning of the foregrounds. We also found that using small chunks also leads to inaccuracies in $C_{HI}$. However, this is not a fundamental issue and would be fixed by simply using more mock maps in our estimates. Trying to keep our models accurate is what motivated our choices for the number of chunks for scales up through $j=6$. For scales $j=7$ and $j=8$, we used 96 chunks as this was the largest value for which our computer was able to compute eigenvectors.

In practice, an analyst using NKL would probably decide on a chunking scheme by trial and error using simulations. However, it may be worth providing some intuition on how these parameters may depend on the survey in question. We should emphasize here that this intuition is really guess work. More precise characterization will need to be determined by trying out NKL in different scenarios.

Suppose one decreased the resolution of one's instrument. This would lead to the angular power spectra like those shown in Fig. \ref{fig:cl_highz} dropping off more quickly as a function of $\ell$. For adequately large (i.e. small $j$) needlet scales, the coefficients would not change much and our approach would stay more or less the same. For smaller needlet scales, noise contamination would become more severe and one may find that larger chunks are required to control errors in $\hat{C}_{FG}$ due to noise leakage.

Now, suppose that one decreased the survey area. Naturally, one would mask out adequately small needlet coefficients and decrease the number of chunks accordingly. It is possible that one could decrease the size of chunks used, as the small and hard to model correlations of distant chunks is now gone. However, this could still pose problems as we expect the anti-correlation between residual foregrounds and signal to scale like $1/N_{p}$.

In the case of an especially small survey area, one may also find that there are not enough non-zero coefficients to adequately estimate $\hat{C}_{FG}$ at lower values of $j$ and in the scaling function. In such a case, one would likely have to increase either the minimum value of $j$ or the value $B$, increasing the number of coefficients per map but providing a coarser partitioning of $\ell$-space.

We found for this high-z case that approximating the foregrounds with DAYENU worked better than with PCA. We believe DAYENU's superior performance in this scenario is likely due to the fact that it treats each pixel in the needlet map individually. On the other hand, PCA gives each pixel the same treatment. This fine-grained approach is useful at higher redshifts where foreground chromaticity varies quite significantly with line of sight.

For the DAYENU foreground approximation, we chose the SVD cutoff to be a factor of $10^{-14}$ below the largest singular value of $\mathbf{C}^{\daleth}$. This value ensured that all foregrounds were captured in the approximation.

The MGNILC cleaning was performed using needlet neighborhoods of size 1500 coefficients. MGNILC was also performed using the same foreground approximation as for NKL.  

For all blind methods, 78 components were used to model the foregrounds. This value was selected by AIC.

\begin{table}
\centering
\begin{tabular}{ccc}
\hline 
\hline
\textbf{Scale}&
\textbf{Number of Chunks}&
\textrm{SNR} \\
\hline
Scaling Function & 3 & 1 \\
$j=4$ & 12 & 4 \\
$j=5$ & 32 & 4\\
$j=6$ & 64 & 4 \\
$j=7$ & 96 & 4 \\ 
$j=8$ & 96 & 4 \\
\hline 
\hline 
\end{tabular}
\caption{A summary of the NKL cleaning parameters used in the high-z test.}
\label{table:nkl_params_highz}
\end{table}

\begin{figure}
    \centering
    \includegraphics[scale=0.5]{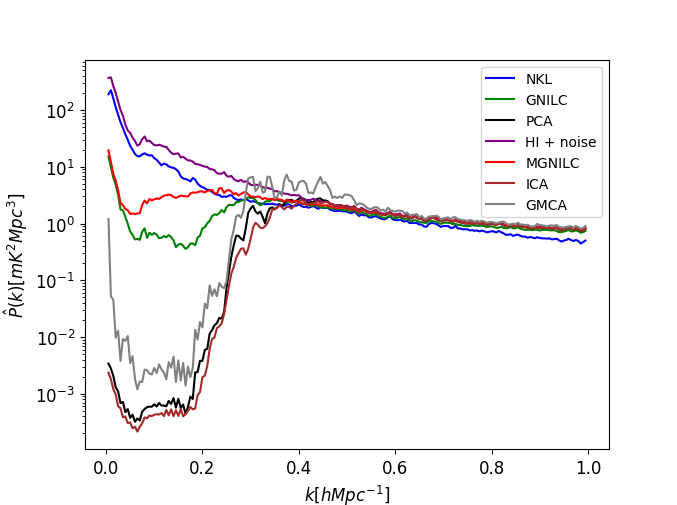}
    \caption{Power spectra estimated from the cleaned high redshift maps. The ``HI + noise" curve shows the power spectrum of the masked, beam-convolved maps. NKL was tested on a second realization of signal plus noise, producing similar results.}
    \label{fig:3dpower_highz}
\end{figure}

In Fig.~\ref{fig:3dpower_highz}, we present the 3D power spectra recovered by the methods described above using equation~(\ref{eq:power}). One may note that these results look more pessimistic than what is seen in other papers on this topic. This is because we are including polarized foregrounds at a higher redshift, which is uncommon. Moreover, most of the sky is kept, including areas with relatively bright foregrounds. One will also notice that all methods tested behaved similarly at small spatial scales ($k \gtrsim 0.6 \, {\rm hMpc}^{-1}$). One exception to this is the NKL curve, which is lower than the others at such scales. This is due to the fact that NKL is the only method tested here that makes a distinction between HI and noise. These small scales carry noise comparable to or greater than the HI signal. As such, NKL looks different than the other methods in this regime. It should be noted that this feature goes away when we set $\mathbf{C}_{S} = \mathbf{C}_{HI} + \mathbf{C}_{N}$. This feature is also not a problem, as the debiased data will show that the HI is still well preserved by NKL at these small scales. At larger spatial scales, there is a clear bifurcation between the blind and non-blind methods. The blind methods clean extremely aggressively at such scales, leaving residuals up to 5 orders of magnitude below the signal plus noise power spectrum. This makes sense, as these large scales will correspond to the foreground-dominated delays such as those seen in Fig.~\ref{fig:delay_comparison}. There is also a notable trough for certain techniques below $k \approx 0.20\,{\rm hMpc^{-1}}$. These scales should correspond roughly to the regime in delay space shown in Fig.~\ref{fig:delay_comparison} where the foregrounds and HI plus noise are approaching the same magnitude.  On the other hand, the non-blind methods seem to be able to preserve some information from within this foreground-dominated region of delay space, having residuals several orders of magnitude closer to that of the true signal plus noise power spectrum. Moreover, NKL seems to provide roughly a factor of 10 to 50 improvement over GNILC at these larger scales. However, it should be noted that all methods incur significant signal loss at the large spatial scales. Even NKL is a factor of a few below the desired level. This is perhaps something that could be compensated for through the use of a transfer function, but we leave such an analysis to future work.  

\begin{figure}
    \centering
    \includegraphics[scale=0.5]{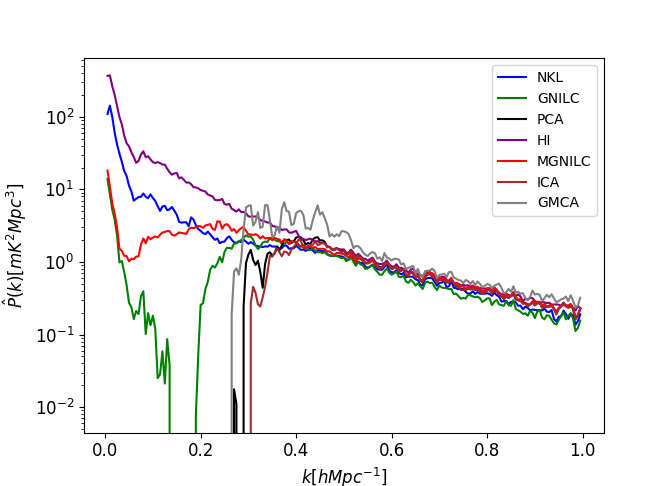}
    \caption{Power spectrum estimates which have been debiased. The ``HI" curve provides the power spectrum of the HI maps after beam convolution and masking.}
    \label{fig:debiased_highz}
\end{figure}

In Fig.~\ref{fig:debiased_highz}, we present power spectrum curves which have been debiased to remove the noise. The power spectra are debiased according to the equation
\begin{equation}
    \hat{P}_{\mathrm{debiased}}(k) = \hat{P}(k) - \hat{P}_{N}(k).
\label{eq:debias}
\end{equation}
In this formula, $\hat{P}(k)$ is the power spectrum estimate obtained from the cleaned maps and $\hat{P}_{N}(k)$ is an estimate of the noise power spectrum obtained from our model of the noise. Note that NKL was debiased slightly differently, with the noise model used being one taking into account the effects of the KL transform on the noise. 
There are a few features in this plot worth noting. First, it should be noted that the debiased power spectra for GNILC, ICA, GMCA and PCA become negative at lower $k$ values. This is due to the loss in both the signal and noise incurred at large scales by these methods, combined with the fact that the noise power spectrum estimate $\hat{P}_{N}$ does not account for this loss. It should be noted that these power spectrum values would likely not become negative had we accounted for this loss. However, even with a more sophisticated debiasing, one would still find these methods being outperformed by MGNILC and NKL. The MGNILC power spectrum estimate never becomes negative. This is because it suffers from less signal loss than the previously mentioned methods. NKL was debiased somewhat differently than the other methods, with the noise power spectrum being estimated from maps cleaned with NKL. It was found that NKL increased the power of the noise at large length scales $k \lesssim 0.35\,\textrm{hMpc}^{-1}$, and decreased the noise power at scales $k \gtrsim 0.35\,\textrm{hMpc}^{-1}$. The use of a modified noise power spectrum resolved these issues in the case of NKL. Moreover, GNILC seems to underestimate the HI power spectrum at higher values of $k$. This is due to the ILC bias described in \citet{delabrouille_2009}. 

\begin{figure*}
    \centering
    \includegraphics[scale=0.5]{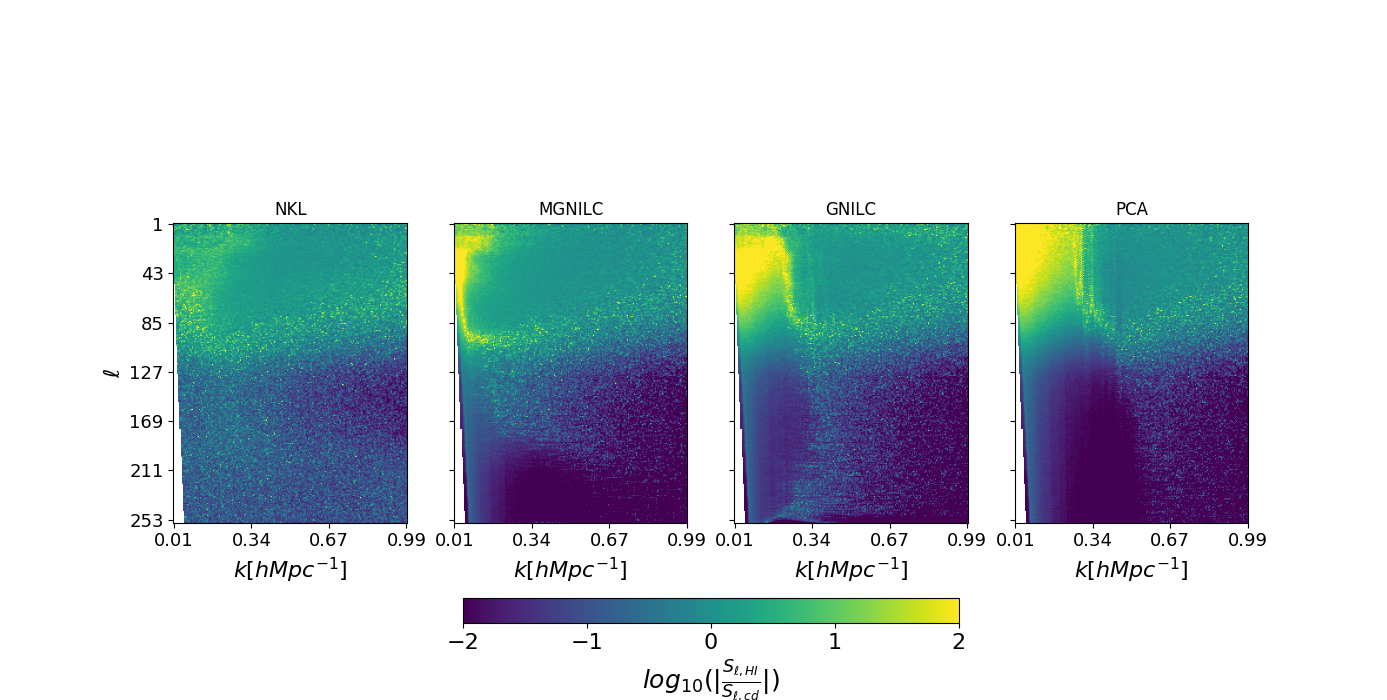}
    \caption{The log-scaled absolute value ratio of the true high-z spherical harmonic power spectrum ($S_{l,HI}$) over the spectrum recovered by cleaning and debiasing ($S_{l,cd}$).}
    \label{fig:Sl_highz}
\end{figure*}

In Fig.~\ref{fig:Sl_highz}, we present graphs illustrating the ratio of $|S_{l,HI}|$ to $|S_{l,cd}|$ as a function of $k$. The variable $S_{l,HI}$ represents the spherical harmonic power spectrum of the HI plus noise. On the other hand, $S_{l,cd}$ represents the spherical harmonic power spectrum obtained from cleaned maps. Note that we also debias $S_{l,cd}$. In this figure, one will notice that there are common features between techniques. First, one will notice larger values of $\log_{10}(|\frac{S_{l,HI}}{S_{l,cd}}|)$ in the top left corner of each plot. This region corresponds to large angular scales and low delays. Since such modes carry the most severe foreground contamination, one will find them being cleaned most aggressively. This effect is most severe for PCA and least severe for NKL. Next, one will notice the plots becoming darker at higher $\ell$. This is due to the fact that the beam convolution applied causes the noise to overpower the HI at small angular scales. The strength of the noise relative to the HI at these scales leads to inaccuracy in the debiasing process. Overall, it appears that NKL has the lowest errors at large spatial scales, as expected from Fig.~\ref{fig:3dpower_highz}.

\begin{figure*} 
    \centering 
    \includegraphics[scale=0.95]{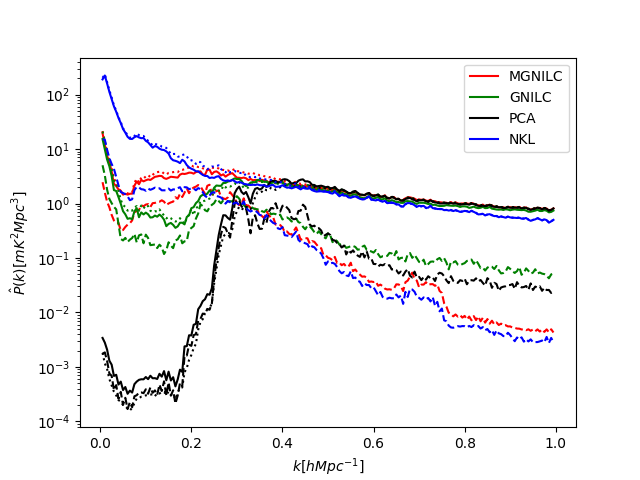}
    \caption{Power spectra illustrating the effects of foreground residuals in the high redshift case. Solid lines correspond to the power spectra of the cleaned maps. Dashed lines correspond to spectra obtained from foreground residuals. Dotted lines correspond to the power spectrum of the HI plus noise present in the cleaned maps.}
    \label{figure:fgresidhz}
\end{figure*}

It is also important to consider residual foregrounds and any bias that these residuals may add to the estimated the HI signal. We will investigate these foreground residuals for GNILC, MGNILC, NKL and PCA. Each of these methods cleans the data by first producing a matrix (or matrices) used for cleaning, followed by application of the matrix (matrices) to the data. For instance, when using PCA, one generates a cleaning matrix $\mathbf{R}$ from a covariance estimate of the data, then cleans the data as $\mathbf{d}_{cleaned} = \mathbf{R}(\mathbf{d}-\overline{\mathbf{d}})$. In order to produce foreground residual estimates, we apply these cleaning matrices to data containing only foregrounds. So, for instance, the foreground residuals for PCA will be 
\begin{equation}
    \mathbf{f}_{residual} = \mathbf{R}(\mathbf{f}-\overline{\mathbf{f}}).
\end{equation}
A similar approach can be taken to obtain the foreground residuals for GNILC, MGNILC and NKL. We also produce an estimate of the HI plus noise present in the cleaned maps in a similar way. In the case of PCA, we would have 
\begin{equation}
    (\mathbf{h}+\mathbf{n})_{cleaned} = \mathbf{R}((\mathbf{h}-\overline{\mathbf{h}})+(\mathbf{n}-\overline{\mathbf{n}})).
\end{equation}
Fig.~\ref{figure:fgresidhz} illustrates the effect that these foreground residuals have on the resulting power spectra. First, notice that the dotted lines, representing the power spectra of the signal plus noise present in the clean maps, tend to be higher than the power spectra of the cleaned maps for the NKL, GNILC, and MGNILC cases. This is indicative of the presence of a negative bias in all three techniques. Moreover, this feature implies that the negative bias present in these techniques decreases the power spectrum estimates, worsening agreement with the true HI power spectrum. The negative bias seen in GNILC and MGNILC is predicted by the ILC bias described in \citet{delabrouille_2009}. The negative bias in NKL was of course predicted in the analysis performed in Section~\ref{subsection:negbias}. 

\begin{figure}
    \centering
    \includegraphics[scale=0.5]{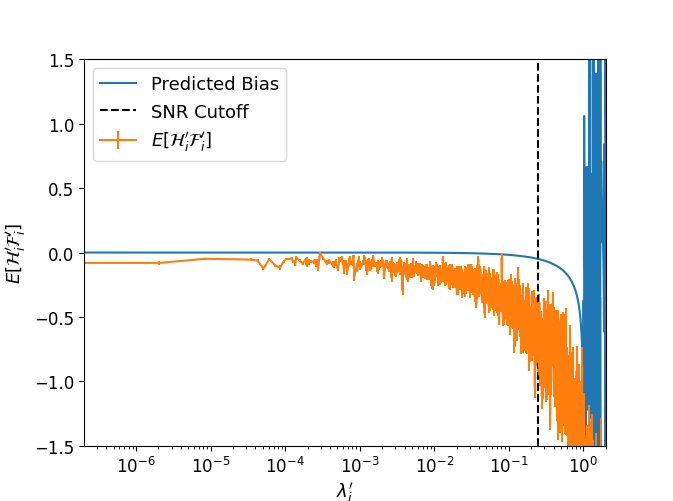}
    \caption{A test of equation~(\ref{eq:fisi}), which modeled the negative bias incurred by NKL, performed on the $j=4$ needlet scale.}
    \label{fig:eq27_test}
\end{figure}

We also explicitly tested equation~(\ref{eq:fisi}) on the results of NKL at needlet scale $j=4$. In this case, we have $N_{p} = 1024$. The results of this are shown in Fig.~\ref{fig:eq27_test}. For this figure, we estimate $E[\mathcal{F}_{i}'\mathcal{H}_{i}']$ by averaging the KL coefficients across all columns of $\underline{\mathbf{h}}$ and $\underline{\mathbf{f}}$. So, we have 
\begin{equation}
    \hat{E}[\mathcal{F}_{i}'\mathcal{H}_{i}'] = \frac{1}{N_{p}} (\phi_{i}'^{T} \underline{\mathbf{h}}) (\underline{\mathbf{f}}^{T}\phi_{i}'). 
\end{equation}
Since we are less interested in foreground-dominated modes, we only present results up to $\lambda = 2$ for clarity. In this plot, one can see that equation~(\ref{eq:fisi}) underestimates the severity of the bias, but is able to roughly capture what is happening. In particular, it predicts that the bias is negative and that it becomes more negative as it one approaches $\lambda' = 1$, the eigenvalue at which the foregrounds and signal plus noise are expected to have equal variance. 

\begin{figure}
    \centering
    \includegraphics[scale=0.5]{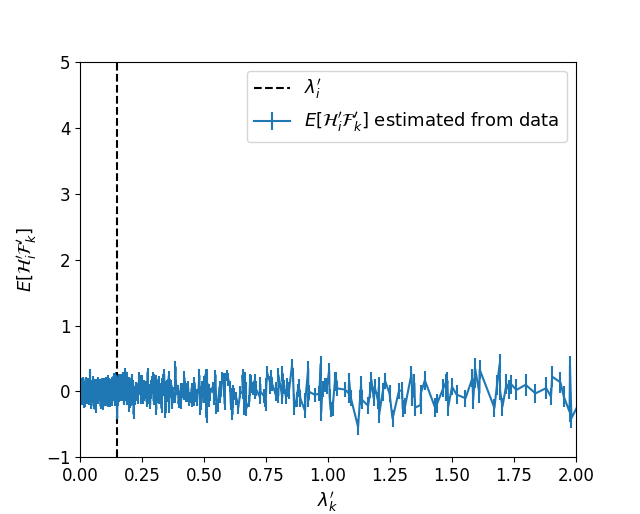}
    \caption{Estimated values of $E[\mathcal{H}_{i}' \mathcal{F}_{k}']$ for $\lambda_{i}'=0.15$.}
    \label{fig:off_diag_test}
\end{figure}

Another aspect of equation~(\ref{eq:fisi}) to consider is the claim that $E[\mathcal{H}_{i}' \mathcal{F}_{k}'] = 0$ for $i \neq k$. We present a test of this in Fig.~\ref{fig:off_diag_test}. In this figure, the blue curve shows estimated values of $E[\mathcal{H}_{i}' \mathcal{F}_{k}']$ for  $\lambda_{i}' = 0.15$. We see that this correlation does not strictly come out to be $0$ as predicted by equation~(\ref{eq:fisi}). Rather, it seems to oscillate about $0$, with $0$ correlation being outside the error bars of many eigenvalues. In this case, the average of these off-diagonal terms (up to $\lambda' = 0.25$) was slightly negative (order $10^{-3}$ and within a standard deviation of $0$. This is much smaller than the diagonal terms  $E[\mathcal{H}_{i}'\mathcal{H}_{i}']$, which are of order 1. As such, these off-diagonal terms will increase errors in the power spectrum estimates, but will not contribute much of a net bias.

\subsection{Test at Lower Redshifts} \label{subsection:lowerz_test}
In this subsection, we test these same foreground removal methods at lower redshifts than in Section~\ref{subsection:highz_test}. In particular, we used 256 evenly spaced frequency channels from 980\,{\rm MHz} up to 1080\,{\rm MHz}. This test was motivated by the fact that the characteristics of the foregrounds change with redshift. In particular, models predict that polarized foregrounds ought to be less severe in this regime. Note that we have again chosen to mask the brightest 15 percent of pixels.  
\begin{figure}
    \centering
    \includegraphics[scale=0.5]{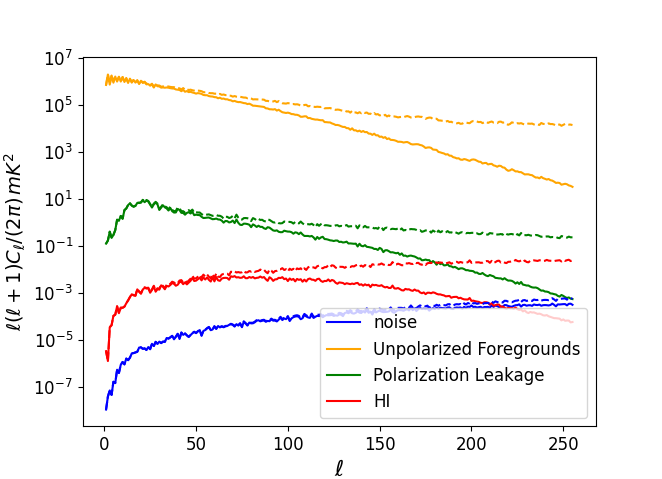}
    \caption{Angular power spectra estimated from unmasked maps using healpy at $1030\,{\rm MHz}$. Solid lines correspond to maps that have undergone beam convolution. Dashed lines correspond to maps that have undergone no beam convolution.}
    \label{fig:cl_lowz}
\end{figure}

In Fig.~\ref{fig:cl_lowz}, we present the angular power spectra estimated at the $1030\,{\rm MHz}$ frequency channel. In this case, for the convolved maps, the HI has a much stronger contribution to the maps than the noise for scales up to $\ell \approx 200$. Although not as severe as in the higher redshift case, the angular resolution of the antenna beams have once again caused significant signal loss at the smaller angular scales. 


For the blind methods, we tried to use AIC to estimate the optimal number of modes to remove. In particular, AIC predicted 4 modes as the best option. However, preliminary checks using the radial power spectrum estimation code from Carucci \footnote{\url{https://github.com/isab3lla/gmca4im}} showed a large spike in the radial power spectrum at low $k_{\nu}$. This spike disappeared when the number of modes was increased to 5, resulting in better agreement between the radial power spectra of the HI plus noise and the cleaned maps. As a result, we chose to remove 5 modes for all blind methods in this test case.

As for GNILC, NKL and MGNILC, we once again used the needlet domain scheme described in Table \ref{table:needlets}. In this case, we again only compute needlets for modes up to $\ell = 255$, as the beams used in this regime have widths $\theta \approx 1^{\circ}$. GNILC was performed in the same way as in Section~\ref{subsection:highz_test}, where we sought to use windows of size $10^{5}$ coefficients. For MGNILC and GNILC, we used neighborhoods of size 3000 coefficients as AIC struggled to select the correct number of modes when using neighborhoods of size 1500. For NKL, we describe the cleaning parameters in Table \ref{table:nkl_params_lowz}.  The number of chunks used has remained more or less the same as in Section~\ref{subsection:highz_test}. We have however, increased the number of chunks for the scaling function and the $j=4$ scale. In addition, note that the same SNR is used for all needlet scales. For this particular test, we found that approximating the foregrounds using the PCA approach was more effective than using DAYENU. That is what we will present in this section. It seems likely that PCA provides an advantage over DAYENU in this scenario for two reasons. For one, the foregrounds in this scenario were quite narrow in delay space. The delay spectra of these foregrounds were particularly narrow compared to those of the higher redshift test case. The fact that the foregrounds vary so rapidly at delays close to $0$ lead to our DAYENU based scheme struggling to accurately estimate the width of the foregrounds. In particular, this approach was usually overestimating the width of the foregrounds in delay space.
On the other hand, the fine-grained nature of the DAYENU-based approach was less important here than in the higher redshift case. This is due to the fact that foreground chromaticity did not vary as much as function of the line of sight in the lower redshift test case.
It should be noted however, that AIC struggled with selecting the correct number of modes during the foreground approximation step. We fixed this simply by
choosing to remove $m+2$ modes, where $m$ was the number of modes selected by AIC. It should also be noted that we chose $\mathbf{C}_{S} = \mathbf{C}_{HI}+\mathbf{C}_{N}$ for this particular test case. This was done since our computer program interpreted $\mathbf{C}_{HI}$ as being singular, preventing the KL transform from being performed. Tacking on $\mathbf{C}_{N}$ was a convenient workaround since the noise is very small in this test and the computations for the $\mathbf{C}_{S} = \mathbf{C}_{N} + \mathbf{C}_{HI}$ test had already been performed. 

\begin{table} 
\centering
\begin{tabular}{ccc}
\hline 
\hline
\textbf{Scale}&
\textbf{Number of Chunks}&
\textbf{SNR} \\
\hline
Scaling Function & 6 & 1 \\
$j=4$ & 12 & 1 \\
$j=5$ & 32 & 1\\
$j=6$ & 64 & 1 \\
$j=7$ & 96 & 1 \\ 
$j=8$ & 96 & 1 \\
\hline 
\hline
\end{tabular}

\caption{A summary of the number of chunks and SNR values chosen for each scale in the low-z test.}
\label{table:nkl_params_lowz}
\end{table}

\begin{figure*} 
    \centering
    \includegraphics[scale=0.75]{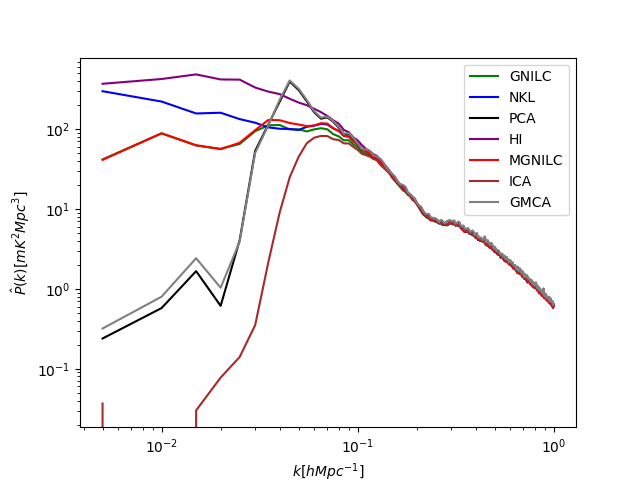}
    \caption{3-dimensional power spectra computed for various techniques in the lower redshift case. The ``HI" curve was estimated from maps that had undergone beam convolution and masking. We also tested NKL on a second realization of the signal plus noise, which produced similar results.}
    \label{fig:3dlowz}
\end{figure*}

\begin{figure*} 
    \centering
    \includegraphics[scale=0.95]{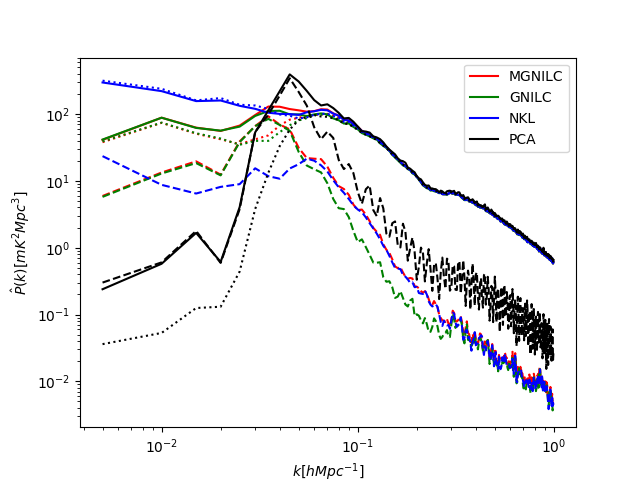}
    \caption{Power spectra illustrating the effects of residual foregrounds for various techniques in the low redshift case. These curves are organized in the same manner as Fig.~\ref{figure:fgresidhz}.}
    \label{fig:fglowz}
\end{figure*}

In Fig.~\ref{fig:3dlowz}, we present power spectra estimated from cleaned maps. It should be noted that the noise is quite low in this test case. As such, the debiasing process of equation~(\ref{eq:debias}) makes an imperceptible difference. Because of this, we only present the debiased results. Note that all methods in this figure were debiased using the same noise power spectrum estimate $\hat{P}_{N}(k)$. In this figure, one will notice first that GMCA and PCA produce essentially identical results, and that they seem to provide the best match to the HI plus noise power spectrum. However, this is somewhat misleading, as the foreground residuals at low $k$ roughly match the HI plus noise power spectrum. This is made clear in Fig. \ref{fig:fglowz}, where one can see that the power spectrum at these scales is mostly due to foreground residuals. The GNILC and MGNILC power spectra also receive a boost from their foreground residuals, but the effect is not as dramatic as in the case of GMCA and PCA. It also appears that the ILC bias in this case is not as severe as in the higher redshift case. This is likely due to the fact that the larger neighborhoods of needlet space were used. As in the higher redshift test case, we find that the presence of foreground residuals decreases the NKL power spectrum at low $k$. However, the effect on NKL is smaller in this case compared to the one at higher redshift.

Interestingly, ICA has performed differently from GMCA and PCA in this context. It appears that ICA has smaller foreground residuals than either GMCA or PCA in this case. Next, one will notice that ICA, MGNILC, GNILC and NKL give very similar results down to $k \approx 0.15\,{\rm hMpc}^{-1}$.  For $k < 0.15\,{\rm hMpc}^{-1}$, the non-blind methods once again provide significantly improved results. For scales $0.03\,{\rm hMpc}^{-1} < k < 0.15\,{\rm hMpc}^{-1}$, we find that NKL, MGNILC and GNILC provide similar results. At the very lowest scale ($k < 0.03\, {\rm hMpc}^{-1}$), we find NKL provides a factor of a few improvement over GNILC and MGNILC. 

Consider again Fig.~\ref{fig:fglowz}, which shows the foreground residuals present in this test case. As mentioned earlier, PCA and GMCA are dominated by foregrounds at $k$ values less than about $0.2\, {\rm hMpc}^{-1}$. For the non-blind methods, we find that foreground residuals are strongest at low values of $k$. This is expected, as such scales will be dominated by contributions from low delays where foregrounds are strongest. However, foreground residuals are not quite as severe as in the higher redshift test case.

\begin{figure*}
    \centering
    \includegraphics[scale=0.5]{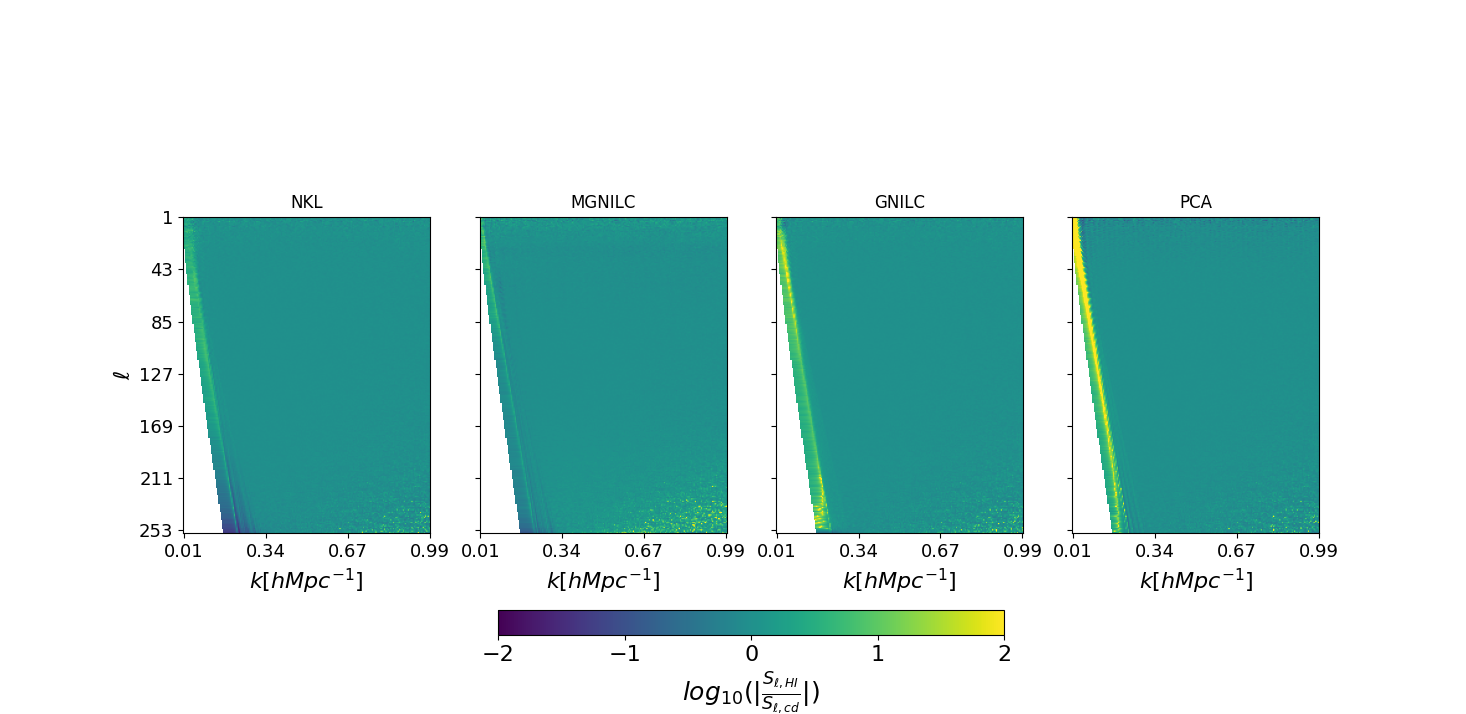}
    \caption{The ratio of the true spherical harmonic power spectrum ($S_{l,HI}$) to the spectrum recovered by cleaning and debiasing ($S_{l,cd}$) for the low-z test.}
    \label{fig:sl_lowz}
\end{figure*}

In Fig.~\ref{fig:sl_lowz}, we show the spherical harmonic power spectra for various cleaning methods. This figure presents the same metric as was presented in Fig.~\ref{fig:Sl_highz}. In this plot, notice first that PCA performs poorly at low $k$ for the lower $\ell$-modes.    At these values of $\ell$, the very lowest $k$ modes were cleaned too aggressively, while the slightly higher $k$ modes were not cleaned enough, leaving behind significant foreground residuals. PCA did not suffer from such under-cleaning at the higher $\ell$-modes, however. For $\ell \gtrsim 200$, MGNILC and NKL provide very similar results, with both leaving behind significant foreground residuals. This is due to the fact that PCA was used to generate the foreground approximation in this test case. When using PCA to generate the foreground approximation, AIC chose to include too few modes. As a result, not all of the foregrounds were included in $\hat{f}$. Any foregrounds not included in $\hat{f}$ will not be cleaned by NKL or MGNILC, leading to the undercleaning observed here. At lower $\ell$, however, it appears that NKL performs best, providing the least signal loss at the lowest $k$ modes. 

\subsection{Performance of NKL Given Modified HI priors}

One source of concern with respect to non-blind methods is how their performance may change subject to systematic effects or modified priors. We will leave systematics testing for future work. For now we will test the performance of NKL subject to modified HI priors.

In \citet{Shaw_2015}, it is mentioned that the CORA software package, which was used for this paper, assumes the HI power spectrum in the flat sky limit is equal to

\begin{equation}
    P_{T_{b}}(\mathbf{k},z,z') = \overline{T_{b}}(z)\overline{T_{b}}(z') (b + f\mu^{2})^{2} D_{+}(z)D_{+}(z')P_{m}(k).
\end{equation}

In this formula, $\overline{T}_{b}$ is proportional to the HI abundance $\Omega_{HI}(z)$ and $b(z)$ is the HI bias.

Although modelling and measurement work has been done, the values of these parameters remain uncertain, particularly at higher redshifts. For the past two decades, much work has been done to derive estimates of $\Omega_{HI}$ from measurements. Figure 14 of \citet{Hu_2019} conveniently provides a summary of $\Omega_{HI}(z)$ estimates out to $z=5$. This figure shows increasing error bars and even disagreement between measurements as redshift increases. This becomes particularly severe for $z \gtrsim 4$.

When searching in the literature for information about the hydrogen bias, one will typically find constraints coming from cross-correlation studies. These cross correlation studies will present estimates of the product $\Omega_{HI}br_{HI}$, where $r_{HI}$ is a correlation coefficient. As a result, most of the information one finds in the literature about $b$ by itself comes from modelling work. The model described in \citet{Castorina_2017} was presented at redshifts $0.8 < z < 5$ and showed significant uncertainties. Possible values of $b(z)$ spanned over a factor of $\approx 2$ at any given redshift. This model however, did predict a smoothly increasing $b(z)$ as a function of redshift.
As a simple way to test the effect of errors in the HI prior, we repeated NKL subject to modified HI priors. In particular, NKL was performed on the same maps as before, but provided with modified estimates of the HI covariance matrices. In these test cases we vary both the bias $b$ and the abundance $\Omega_{HI}$ assumed by the prior. In particular, we scale $\Omega_{HI}$ by a constant factor and change the bias from a constant to a linear function. The coefficients of this linear curve are chosen to ensure $b(z_{min}) = 1$ and $b(z_{max}) = 2$. To summarize, we modify the priors in three different ways:
\begin{enumerate}
    \item $\Omega_{HI} \rightarrow \Omega_{HI}/2$ \textrm{and} $b(z) = 1 \rightarrow b(z) = az + c$
    \item $\Omega_{HI} \rightarrow 2\Omega_{HI}$ \textrm{and} $b(z) = 1 \rightarrow b(z) = az + c$
    \item $\Omega_{HI} \rightarrow \Omega_{HI}/2$.
\end{enumerate}

These robustness tests were performed both in the higher redshift scenario and the lower redshift scenario. The coefficients chosen for the bias were adjusted in each case to ensure that the bias varied from 1 to 2 over the redshift range.

\begin{figure}
    \centering
    \includegraphics[scale=0.49]{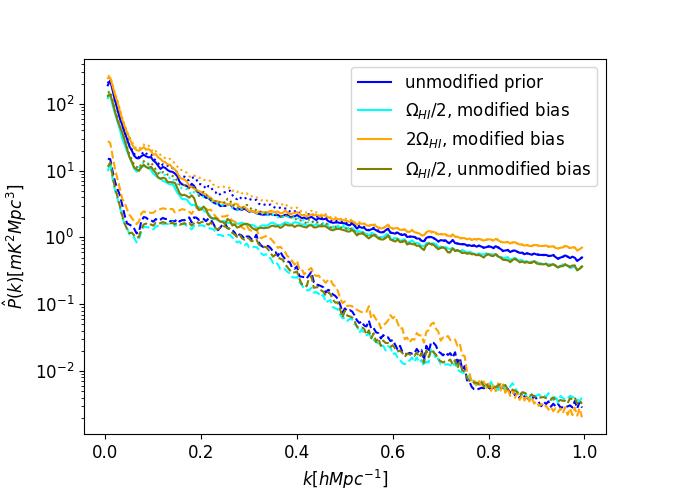}
    \caption{A summary of NKL performance subject to modified priors in the higher redshift scenario. The curves shown here should be interpreted in the same as Fig.~\ref{figure:fgresidhz}.}
    \label{fig:summary_highz_robust}
\end{figure}

In Fig.~\ref{fig:summary_highz_robust}, we present results from the robustness check at higher redshifts. As can be seen, NKL is fairly robust to errors in the prior in this scenario. All priors have provided roughly similar levels of foreground residuals. One will also notice that the $\Omega_{HI}/2$ curves are lower and the  $2\Omega_{HI}$ curve is higher. This is due to the fact that the KL step of NKL chooses the number of modes to discard based on the expected signal to ratio. This SNR estimate is of course dependent upon the assumed HI amplitude. It should also be noted here that the two $\Omega_{HI}/2$ curves look quite similar. This would imply that errors in the bias prior did not make much of a difference in this case.

\begin{figure}
    \centering
    \includegraphics[scale=0.49]{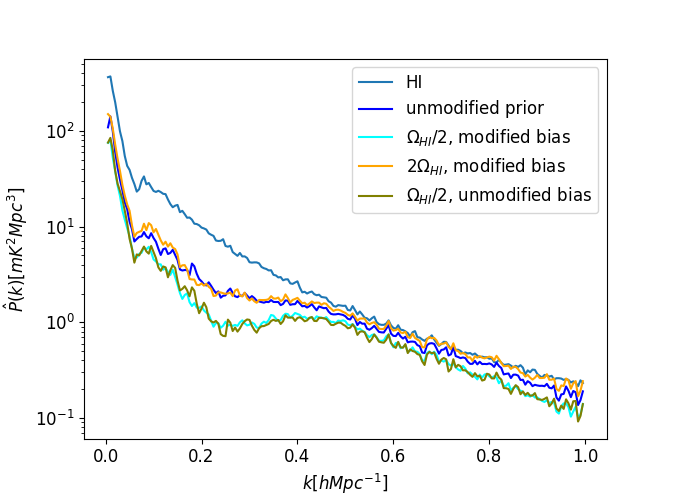}
    \caption{Debiased NKL power spectra in the high redshift test case, subject to modified priors.}
    \label{fig:highz_robust_debiased}
\end{figure}

In Fig.~\ref{fig:highz_robust_debiased}, we present debiased power spectra for the high redshift test case. As one could predict from Fig.~\ref{fig:summary_highz_robust}, we see that the $2\Omega_{HI}$ provides roughly similar performance to the correct prior case upon debiasing. On the other hand, the cases with $\Omega_{HI}/2$ suffer from some additional signal loss.

\begin{figure}
    \centering
    \includegraphics[scale=0.49]{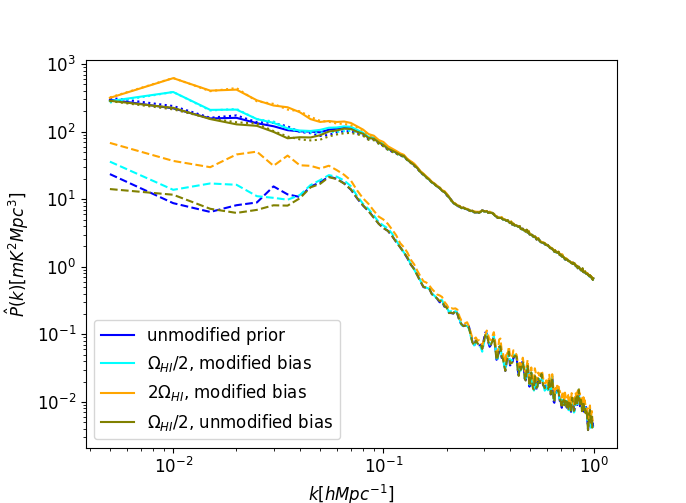}
    \caption{A summary of NKL performance subject to modified priors in the lower redshift scenario. The curves shown here should be interpreted in the same way as Fig.~\ref{figure:fgresidhz}.}
    \label{fig:summary_lowz}
\end{figure}

In Fig.~\ref{fig:summary_lowz}, we present a summary of results with varied priors in the low-z case. As in the higher redshift scenario, decreasing $\Omega_{HI}$ without adjusting the bias leads to roughly similar results. However, adjusting the bias seems to have more of an impact, with both the $\Omega_{HI}/2$ and $2\Omega_{HI}$ test cases showing larger power spectrum values at low $k$ compared to the case with correct priors. Further testing found that NKL discarded the same number of modes in both the correct prior and $\Omega_{HI}/2$ with linear bias test cases. This would imply that the performance difference between the cyan and blue curves of Fig.~\ref{fig:summary_lowz} is due to differences in the eigenmodes generated for the KL transform, rather than NKL discarding fewer modes due to the change in prior.  

\begin{figure}
    \centering
    \includegraphics[scale=0.5]{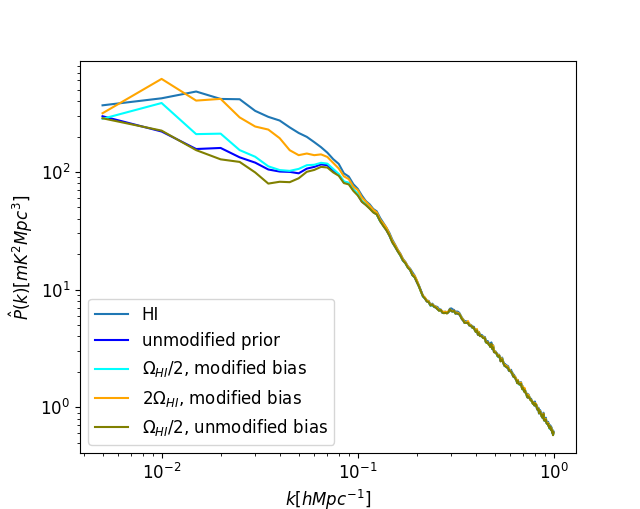}
    \caption{Debiased NKL power spectra in the lower redshift test case, subject to modified priors.}
    \label{fig:robust_lowz}
\end{figure}

In Fig.~\ref{fig:robust_lowz}, we present debiased power spectra for the lower redshift scenario. In this figure, one sees that decreasing $\Omega_{HI}$ without adjusting the bias leads to similar results, with the cleaning being conducted slightly more aggressively. The test cases with modified bias assumptions oddly lead to better agreement, with some over estimation at the lowest values of $k$ in the $2\Omega_{HI}$ case.  

\section{Discussion and Conclusion} \label{section:conclusion}
In this paper we introduced the NKL and MGNILC techniques for cleaning foregrounds from HI intensity maps. Moreover we tested these methods, and various others, on full sky maps in two different bands of observations ($[980\,\mathrm{MHz},1080\,\mathrm{MHz}]$ and $[400\,\mathrm{MHz},500\,\mathrm{MHz}]$). These tests were conducted assuming a hypothetical telescope similar to MeerKAT operating in single-dish mode. Instrumental effects were modelled simply, with noise being drawn from a Gaussian distribution and beam effects modeled by convolution with Gaussians of frequency-dependent width. Moreover, we assumed full sky coverage and masked out the brightest 15 percent of pixels.

At higher redshifts, where foregrounds are more severe, we found that non-blind methods such as GNILC and NKL outperform blind methods by several orders of magnitude at large spatial scales. However, all methods suffered significant signal loss. The most accurate method was NKL, which was still off by an order of magnitude in the higher redshift test and a factor of 2 or 3 in the lower redshift case. It may be possible to compensate for such signal loss using a transfer function method. For instance, the work presented in \citet{Cunnington_transfer} seems to show promising results when signal loss is at the 50 percent level. 

We also tested the robustness of NKL against modified priors. We found that the performance of NKL remains fairly stable while varying $\Omega_{HI}$ and the bias $b$. It was found that varying $\Omega_{HI}$ changed how aggressively NKL cleaned, while varying the bias lead to different behavior at low $k$ in the lower redshift scenario. However, even when provided with modified priors, NKL still provided better performance at large length scales than the other map-space cleaning methods described in this paper.

We must emphasize that these tests do not tell the entire story of foreground removal. For instance, convolution with a Gaussian beam is not quite realistic. A more realistic beam would be more complicated than this, including sidelobes and asymmetries. Moreover, a realistic mapmaking process is not so simple as just convolving with a beam. One has to perform some sort of maximum likelihood estimate based on the raw data coming from the antennas. Such a process may lead to mapmaking artifacts that aren't so easily described. In addition, one must also consider inevitable imperfections in beam calibration. Such imperfections would likely provide their own artifacts and possibly add additional chromaticity to the foregrounds.  We plan to address these issues in future work.

For perspective, we would like to recall the visibility-based work presented in \citet{Shaw_2015}. In that paper, polarized foregrounds were removed by projecting the visibilities onto the null-space of the polarized beam matrix. A KL transform was then used to clean the unpolarized foregrounds from the visibilities. That approach recovered the HI signal effectively down to $k \approx 0.02\, {\rm hMpc}^{-1}$ for a simplified version of the CHIME telescope operating between 400 and 500\,{\rm MHz}. One conclusion, however, was that the main beam widths needed to be understood to an accuracy of $0.1 \%$ for this method to work. 

In this paper, our motivation for further investigation of map-based techniques was that they may be more robust to beam mis-calibration than are visibility-based methods. For instance, Shaw's methods would not work in a situation where beamwidth errors are at the  $1 \%$ level. This leads to an incorrect model of $C_{FG}$, preventing the Karhunen-Loève transform from being effective. We mention $C_{FG}$ in specific because it is more sensitive to errors than $C_{HI}$. Having $1 \%$ errors in $C_{FG}$ can be hazardous when dealing with such a large dynamic range as in HI intensity mapping. However, NKL and MGNILC estimate $C_{FG}$ directly from the data, hopefully providing it with some robustness against calibration issues.  

As a next step, we intend to use the driftscan package, which employs the m-mode formalism of \citet{Shaw_2015}, to produce more realistic maps. This approach creates visibilities from a beam model. We can then use a modified version of the package to generate Wiener-filtered maps from these visibilities. We will then introduce beam-width errors to test how robust these methods are. We also intend to introduce a method similar to NKL which acts in the visibility domain. Such a method may be robust to errors in the beam widths, as the foreground covariance would be estimated from the data and not from a beam model.

\section*{Acknowledgements}

We thank Le Zhang, Ted Bunn, Calvin Osinga and our anonymous reviewer for providing helpful feedback on this work. This work was partially supported by NSF Award AST-1616554, the University of Wisconsin Graduate School, the Thomas G. Rosenmeyer Cosmology Fund, and by a student award from the Wisconsin Space Grant Consortium.

\section*{Data Availability}

The data underlying this article will be shared on reasonable request to the corresponding author.



\bibliographystyle{mnras}
\bibliography{example} 




\appendix

\section{Details of the Error Analysis}
In this appendix, we provide derivations of the expressions provided in Section \ref{subsection:negbias}. We derive the formulas in this subsection using first order perturbation theory. To begin, let's consider the generalized Eigenvalue problem at hand. It turns out that equation~(\ref{pert_eig}) can be massaged into 
\begin{equation}
    \mathbf{C}_{S}^{-1/2} \hat{\mathbf{C}}_{FG} \mathbf{C}_{S}^{-1/2} \phi_{i}'' = \lambda_{i} \phi_{i}''. 
\end{equation}
In this formula, $\phi_{i}'' = \mathbf{C}_{S}^{1/2}\phi_{i}'$ and $\mathbf{C}_{S}^{1/2}$ is the Hermitian square root of $\mathbf{C}_{S}$. Note that this is the same as the equation we would get had we set up the generalized Eigenvalue equation using variables whitened with $\mathbf{C}_{S}^{-1/2}$. 

It turns out that the math involved in this problem is easier when whitened variables are used. As such, any symbols used during the rest of this derivation will be used to represent whitened variables. So, for instance, $f_{p}$ will refer to row $p$ of $\mathbf{C}_{S}^{-1/2} \underline{\mathbf{f}}$. Moreover, $\phi_{i}'$ will refer to the whitened eigenvector $\phi_{i}''$. This will not change our results, since it can be easily shown that $E[\mathcal{F}_{i}'\mathcal{H}_{k}']$ does not change upon whitening of the variables. 

To first order, the perturbed eigenvector $\phi_{i}'$ will look like 

\begin{equation} \label{eq:pertvec}
    \phi_{i}' = \phi_{i} + \sum_{j \neq i} \frac{\phi_{j}^{T} \mathbf{\Delta} \phi_{i} }{\lambda_{i}-\lambda_{j}} \phi_{j}.
\end{equation}
In this formula, 
\begin{equation} \label{eq:errors}
    \mathbf{\Delta} = \mathbf{\Delta} _{FG} + \mathbf{C}_{h'h'} + 
    \mathbf{C}_{fh'} + \mathbf{C}_{n'h'} +  \mathbf{C}_{n'n'} +  \mathbf{C}_{fn'}.
\end{equation} 
Refer to Section \ref{subsection:negbias} for definitions of the terms on the right of equation~(\ref{eq:errors}).
Let's now consider the correlation between residual foregrounds and HI. We will find that 

\begin{equation}
    E[\mathcal{F}_{i}'\mathcal{H}_{k}'] = E[(\phi_{i} + \Delta \phi_{i})^{T} f_{p} h_{p}^{T}(\phi_{k} + \Delta \phi_{k})].
\end{equation}
Refer to Section \ref{subsection:negbias} for definitions of $\mathcal{F}_{i}'$ and $\mathcal{H}_{i}'$. Note also that $a_{p}$ refers to column $p$ of matrix $a$. This expectation value ought to be equal to 0 in the unperturbed case. We will also discard the second order term, giving 

\begin{equation}
    E[\Delta \phi_{i}^{T} f_{p} h_{p}^{T}\phi_{k}] +  E[\phi_{i}^{T} f_{p} h_{p}^{T}\Delta\phi_{k}].
\end{equation}
Recall that $\Delta \phi_{x}$ really consists of several contributions
\begin{multline}
\Delta\phi_{x} = \\ \sum_{j \neq x} \frac{\phi_{j}^{T} (\mathbf{\Delta} _{FG} + \mathbf{C}_{h'h'} + 
    \mathbf{C}_{fh'} + \mathbf{C}_{n'h'} +  \mathbf{C}_{n'n'} +  \mathbf{C}_{fn'})\phi_{x} }{\lambda_{x}-\lambda_{j}} \phi_{j}.
\end{multline}
The term involving $\Delta_{FG}$ will not contribute to $E[\mathcal{F}_{k}'\mathcal{H}_{i}']$ as it will consist of the expectation of products of 3 terms depending on the foregrounds and 1 term depending on the HI. Since the HI and foregrounds are uncorrelated, this should result in an expectation value of 0. Similar arguments can be made for $\mathbf{C}_{h'h'}$, $\mathbf{C}_{n'h'}$, $\mathbf{C}_{n'n'}$ and  $\mathbf{C}_{fn'}$, which will each involve odd degrees of foregrounds, noise or HI, resulting in 0 (or at least small) expectation. Small expectation is mentioned since terms like $h'_{p}$ and $f_{p}$ will be correlated in principle, since $h'_{p}$ does depend on the foregrounds. However, we will assume that such correlation are small.   

It will be the case however, that $\mathbf{C}_{fh'}$ will create a non-negligible correlation between the foreground residuals and HI. We will have
\begin{multline} \label{eq:crosscorr}
E[\mathcal{F}_{i}'\mathcal{H}_{k}'] = \sum_{j \neq i} \frac{1}{\lambda_{i}-\lambda_{j}} (E[\phi_{j}^{T} \mathbf{C}_{fh'} \phi_{i} \phi_{j}^{T} f_{p} h_{p}^{T} \phi_{k}] \\ + \sum_{j \neq k} \frac{1}{\lambda_{k}-\lambda_{j}} E[\phi_{i}^{T} f_{p} h_{p}^{T} \phi_{j} \phi_{j}^{T} \mathbf{C}_{fh'} \phi_{k}]).
\end{multline}
Let's start by considering the first term of the right side of equation~(\ref{eq:crosscorr}). Applying the definition of $\mathbf{C}_{fh'}$, we obtain 
\begin{equation} \label{eq:term1}
    \sum_{j \neq i} \frac{1}{\lambda_{i}-\lambda_{j}} \sum_{q} \frac{1}{N_{p}} E[\phi_{j}^{T} (h_{q}'f_{q}^{T} + f_{q}h_{q}'^{T}) \phi_{i} \phi_{j}^{T} f_{p} h_{p}^{T} \phi_{k}].
\end{equation}
For first term of equation~(\ref{eq:term1}), we will have
\begin{equation} \label{eq:shouldbezero}
\Sigma_{j \neq i} \frac{1}{\lambda_{i}-\lambda_{j}} \Sigma_{q} \frac{1}{N_{p}}  E[\phi_{j}^{T} h'_{q}f_{q}^{T}\phi_{i} \phi_{j}^{T} f_{p} h_{p}^{T} \phi_{k}].
\end{equation}
Let us now assume that $h'$ and $f$ are uncorrelated. This assumption is not strictly true, as the filter used to generate $h'$ will be a function of the foregrounds. However, we found in numerical tests that correlations between $h'$ and $f$ were small. So, this approximation is safe to make. Next, we rearrange terms inside the expectation value and using our recent assumption, we can express equation~(\ref{eq:shouldbezero}) as  
\begin{equation} \label{eq:sumofprods}
\sum_{j \neq i} \frac{1}{\lambda_{i}-\lambda_{j}} \sum_{q} \frac{1}{N_{p}}  E[\phi_{j}^{T}h_{q}'h_{p}^{T} \phi_{k}] E[\phi_{i}^{T} f_{q} f_{p}^{T} \phi_{j}].
\end{equation}
Next, note that the rapid angular variation of the signal implies that terms of in the sum of equation~(\ref{eq:sumofprods}) will be suppressed when $q$ differs from $p$. As such, we will approximate equation~(\ref{eq:sumofprods}) as 
\begin{equation}
\Sigma_{j \neq i} \frac{1}{\lambda_{i}-\lambda_{j}} \frac{1}{N_{p}}  E[\phi_{j}^{T}h_{p}'h_{p}^{T} \phi_{k}] E[\phi_{i}^{T} f_{p} f_{p}^{T} \phi_{j}].
\end{equation}
Assuming that $E[f_{p}f_{p}^{T}]$ varies little as a function of $p$, then we can say $E[\phi_{i}^{T} f_{p} f_{p}^{T} \phi_{j}] \approx 0$. This assumption should be a good one when chunk sizes are small enough that the foreground temperature varies little within each chunk. On other hand, consider the case where one has a single chunk covering the whole map. In this case, the columns of $f$ will correspond to individual lines of sight. In such a case, $E[f_{p}f_{p}^{T}]$ will vary significantly with $p$ and our assumption is no longer valid. 

Thus, we find that the first term on the right side of equation~(\ref{eq:term1}) is approximately 0, as long as the chunk sizes are chosen to be adequately small. 
As for the second term of equation~(\ref{eq:term1}), we can do some rearranging and apply our previous assumptions to say that 
\begin{multline}
\sum_{j \neq i} \frac{1}{\lambda_{i}-\lambda_{j}}  \sum_{q} \frac{1}{N_{p}} E[\phi_{j}^{T} 
 f_{q}h_{q}'^{T} \phi_{i} \phi_{j}^{T} f_{p} h_{p}^{T} \phi_{k}] \\ 
\approx \frac{1}{N_{p}} \Sigma_{j \neq i} \frac{1}{\lambda_{i}-\lambda_{j}} E[\phi_{j}^{T} f_{p}f_{p}^{T}\phi_{j}] E[\phi_{i}^{T}h_{p}' h_{p}^{T} \phi_{k}].
\end{multline}
Let's define the matrix $\alpha_{ik} = E[\phi_{i}^{T}h_{p}' h_{p}^{T} \phi_{k}]$. In a numerical test, we found that $\alpha_{ik}$ roughly follows the relation 
\begin{equation} \label{eq:alphaapprox}
\alpha_{ik} \approx \begin{cases} 
      0 & \lambda_{i} < 1 \\
      \delta_{ik} & \lambda_{i} > 1.  
   \end{cases}
\end{equation}
This makes sense, as $h_{p}'$ is just a version of $h_{p}$ that has been smoothed using some filter (either PCA or DAYENU). We know that modes with large $\lambda_{i}$ will couple strongly to the foregrounds, and thus to the smooth parts of $h_{p}$. As such, it should be the case that 
\begin{equation}
    E[\phi_{i}^{T}\overline{h}_{p} h_{p}^{T} \phi_{k}] \approx E[\phi_{i}^{T}h_{p} h_{p}^{T} \phi_{k}] = \delta_{ik}
\end{equation}
when $\lambda_{i}$ is large. On the other hand, for smaller $\lambda_{i}$, one would expect that the mode would couple most strongly to higher delay parts of $h_{p}$. Since $h_{p}'$ has been smoothed, it should be the case that 
\begin{equation}
    |E[\phi_{i}^{T}h_{p}' h_{p}^{T} \phi_{k}]| << 1
\end{equation}
for modes with small $\lambda_{i}$. Using this result, we obtain
\begin{multline}
    \sum_{j \neq i} \frac{1}{\lambda_{i}-\lambda_{j}}  \sum_{q} \frac{1}{N_{p}} E[\phi_{j}^{T} 
 f_{q}h_{q}'^{T} \phi_{i} \phi_{j}^{T} f_{p} h_{p}^{T} \phi_{i}] \\ \approx \frac{\delta_{ik}}{N_{p}} \sum_{j \neq i} \frac{\alpha_{ii} \lambda_{j}}{\lambda_{i}-\lambda_{j}}. 
\end{multline}
We can go through the same process for the other term involved in equation~(\ref{eq:crosscorr}). This results in
\begin{equation}
  \frac{\delta_{ik}}{N_{p}}\Sigma_{j \neq i} \frac{\alpha_{jj}\lambda_{i}}{\lambda_{i}-\lambda_{j}},
\end{equation}
Where $\delta_{ik}$ is a Kronecker delta function. Combining these two terms, we find 
\begin{equation}
    E[\mathcal{F}_{i}'\mathcal{H}'_{k}] \approx \frac{\delta_{ik}}{N_{p}} \Sigma_{j \neq i} \frac{(\alpha_{ii} \lambda_{j} + \alpha_{jj}\lambda_{i})}{\lambda_{i}-\lambda_{j}}.
\end{equation}


\bsp	
\label{lastpage}
\end{document}